\documentclass[11pt,onecolumn,amssymb,nofootinbib]{revtex4}
\usepackage{amsmath, amsthm, amscd, amssymb}
\usepackage{bm}
\usepackage{bbm}

\begin{document}

\title{\bf Classical Gauged Massless Rarita-Schwinger Fields}

\author{Stephen L. Adler}
\email{adler@ias.edu} \affiliation{Institute for Advanced Study,
Einstein Drive, Princeton, NJ 08540, USA.}

\begin{abstract}
We show that, in contrast to known results in the massive case, a minimally gauged
massless Rarita-Schwinger field yields a consistent classical theory, with a
generalized fermionic gauge invariance realized as a canonical transformation.
To simplify the algebra, we study a two-component left chiral reduction of the massless theory.
We formulate the classical theory in both Lagrangian and Hamiltonian form for a general
non-Abelian gauging, and analyze the constraints and the Rarita-Schwinger gauge invariance of the action.
An explicit wave front calculation for
Abelian gauge fields shows that wave-like modes do not propagate with superluminal velocities.  An analysis of Rarita-Schwinger spinor
scattering from gauge fields shows that adiabatic decoupling fails in the limit of zero gauge field amplitude, invalidating various
``no-go'' theorems based on ``on-shell'' methods that claim to show the impossibility of gauging Rarita-Schwinger fields.  Quantization
of Rarita-Schwinger fields, using many formulas from this paper, is taken up in the following paper.
\end{abstract}

\maketitle

\section{Introduction}

\subsection{Motivations and Background}
Cancelation of gauge anomalies is a basic requirement for constructing grand unified models, and the usual
assumption is that anomalies must cancel among spin $\frac{1}{2}$ fermion fields.  However,  a 1985 paper
of Marcus \cite{marcus} showed that in principle an $SU(8)$ gauge theory can be constructed with
spin $\frac{3}{2}$ Rarita-Schwinger fermions playing a role in anomaly cancelation, and we have recently
constructed \cite{adler} a family unification model incorporating this observation.  Using gauged spin $\frac{3}{2}$ fields in a grand unification model raises the question of whether such fields admit a consistent quantum, or even classical
theory. It is well known, from papers of Johnson and Sudarshan \cite{johnson} and Velo and Zwanziger \cite{velo}, and
much subsequent literature (see e.g. Hortacsu \cite{hortacsu} and  Deser and Waldron \cite{deser}),  that theories of massive gauged Rarita-Schwinger fields have serious problems. Does setting
the fermion mass to zero eliminate these difficulties?

The lesson we have learned from the success of the Standard Model is that fundamental fermion masses lead to
problems and are to be avoided; all mass is generated by spontaneous symmetry breaking, either through
coupling to the Higgs boson or through the formation of  chiral symmetry breaking fermion condensates.  So from a modern point
of view, the Rarita-Schwinger theory with an explicit mass term is suspect. Several  hints that the behavior of the massless
theory may be satisfactory are already apparent from a study of the zero mass limit of formulas in the Velo--Zwanziger paper.  First, in their demonstration of superluminal
signaling, the problematic sign change that they find for large $B$ fields \big(Eq. (2.15) of \cite{velo}\big) is not present when the mass is set to zero.
Second, when the mass is zero, the secondary constraint that they derive \big(Eq. (2.10) of \cite{velo}\big) appears as a factor in the change in the action  under a Rarita-Schwinger gauging $\delta \psi_{\mu} = D_{\mu} \epsilon$, with $D_{\mu}$ the usual gauge covariant derivative. \big(This statement is not in \cite{velo}, but is an easy calculation from their Eqs. (2.1)--(2.3), with the $D_{\mu}$ of this paper their $-i\pi_{\mu}$.\big)  Hence, the constrained action in the massless gauged Rarita-Schwinger  theory has a fermionic gauge invariance that is the natural generalization of the fermionic
gauge invariance of the massless free Rarita-Schwinger theory.  Third, their formula for the anticommutator \big(Eq. (4.12) of
\cite{velo}\big)  in the zero mass case
develops an apparent singularity  in the limit of vanishing gauge field $B$, and so their quantization does not limit to the standard free
theory quantization.  However,  since the massive theory does not have a fermionic gauge invariance, Ref.
\cite{velo} does not include a gauge-fixing term analogous to that used in the massless case, but gauge fixing is needed
to get a consistent quantum theory for a free massless Rarita-Schwinger field.  So these observations, following from the
equations in \cite{velo}, suggest that a study of the massless Rarita-Schwinger field coupled to spin-1 gauge fields is in order.

In a different and more recent setting, massless Rarita-Schwinger fields appear consistently coupled to gravity as the gravitinos of supergravity, as discussed by
Das and Freedman \cite{dasa}.  Grisaru, Pendleton  and van Nieuwenhuizen \cite{grisaru}  have shown that soft spin $\frac{3}{2}$ fermions {\it must} be coupled to
gravity as in supergravity, in an analysis based on the free particle external line pole structure of spin $\frac{3}{2}$ fields that do not have spin 1 gauge couplings. Their result has been extended to  gauged spin $\frac{3}{2}$ fields  in various recent ``no-go'' theorems based on
 ``on-shell'' methods \cite{porrati}, \cite{mcgady}, that again assume a free particle external line pole structure.  None of these papers have analyzed the gauged
Rarita-Schwinger equation to determine the asymptotic field structure.   Thus, these papers
 do not prove that there cannot be a consistent theory of
massless, gauged Rarita-Schwinger fields, so again a detailed study of this possibility is warranted.

\subsection{Outline of the paper, and summary}
With these motivations and background in mind, we embark in this paper on a detailed study of the classical theory of a minimally gauged massless Rarita-Schwinger field.  In Sec. 2, we give the Lorentz covariant Lagrangian for a gauged four-component Rarita-Schwinger spinor field, derive the source current for the gauge field, and check that it is gauge-covariantly conserved.  We also give the Lorentz covariant form of the constraints,  of the fermionic gauge transformation, and of the symmetric stress-energy tensor, and  briefly discuss the generalization to
non-flat metrics.    Since in the massless
case left chiral and right chiral components of the field decouple,  in Sec. 3 we rewrite the Lagrangian for left
chiral components in terms of two-component spinors and Pauli matrices, which simplifies the subsequent analysis.  We then give
the Euler-Lagrange equations in two-component form, and use them to analyze the structure of constraints and the fermionic gauge transformation of the action. In Sec. 4 we introduce canonical momenta for the Rarita-Schwinger field components, which are used to define classical Poisson brackets,  and discuss the role of the constraints as generators of gauge transformations under the bracket operation.  We  show that the constraints group into two sets of four, within each of which there are vanishing Poisson brackets. In Sec. 5, we argue that fermionic
gauge transformations give a generalized form of gauge invariance, corresponding to the presence of redundant gauge degrees of freedom,
by studying the properties of both infinitesimal and general finite gauge transformations. We show that infinitesimal gauge transformations are an 
 invariance of the constrained action functional that governs the influence of Rarita-Schwinger fields on gauge and gravitational fields.  We show that finite gauge transformations take
the form of generalized auxiliary fields, which lead to an extended action that has an exact invariance under fermionic gauge transformations.
In Sec. 6 we specialize to the case of an Abelian gauge field \big(as in \cite{velo}\big) and analyze the wavefront structure, showing that physical wave modes  propagate with luminal velocities; an extension of this discussion, showing that gauge modes are subluminal is given in Appendix B.
In Sec. 7, making a transition to first quantization,  we analyze Rarita-Schwinger fermion scattering from an
 Abelian gauge potential.  We show that the asymptotic state structure assumed in ``on-shell no-go'' theorems is not realized, but
 that a consistent scattering amplitude can be formulated using an analog of the distorted wave Born approximation.   In Sec. 8  we give a
 brief summary and discussion, and in Appendix A we summarize our notational conventions and some useful identities.  We suggest that the reader skim through Appendix  A before going on to Sec. 2, since things stated in Appendix A are not repeated in the body of the paper.  In the paper
 that follows this one we build on our analysis to discuss quantized Rarita-Schwinger fields.

\section{Lagrangian and covariant current conservation in four-component form}
\subsection{Flat spacetime}
The action for the massless Rarita-Schwinger theory is
\begin{align}\label{eq:action}
S(\psi_{\mu}) = &\frac{1}{2} \int d^4x \,\overline{\psi}_{\mu \alpha u} R^{\mu \alpha u}~~~,\cr
 R^{\mu\alpha u}=&i\epsilon^{\mu\eta\nu\rho}(\gamma_5\gamma_{\eta})^{\alpha}_{~\beta}(D_{\nu}\psi_{\rho}^{\beta})^u~~~,\cr
 (D_{\nu}\psi_{\rho}^{\beta})^u\equiv&\partial_{\nu}\psi_{\rho}^{\beta u}+gA_{\nu v}^{u} \psi_{\rho}^{\beta v}~~~,\cr
 A_{\nu v}^u=&A_{\nu}^A t_{A v}^u~~~,\cr
\end{align}
with $\psi^{\mu \alpha u}=\psi^{\mu \alpha u}(\vec x,t)$ a four-vector four-component spinor, with four-vector index $\mu=0,...,3$, spinor index $\alpha=1,...,4$, and $SU(n)$ internal
symmetry index $u=1,...,n$, with $SU(n)$ gauge generators $t_A,\,A=1,...,n^2-1$.  Taking  $u$ to range from 1 to  $n$ means that, for definiteness, we are
 assuming that the spinors transform according to the fundamental representation of the $SU(n)$ internal symmetry group, but other representations and other compact Lie groups can be
 accommodated by assigning the  internal indices $u$ and $A$ the appropriate range.  Note that $t_A$, $ A_{\nu v}^u$, and $D_{\nu}$ all commute with the gamma matrices and  the Pauli
spin matrices from which the gamma matrices are constructed, and for an Abelian internal symmetry group, the indices $u$ and $A$ are not needed.  Using
\begin{equation}
 \overline{\psi}_{\mu\alpha u}=\psi_{\mu\beta u}^{\dagger}i(\gamma^0)^{\beta}_{~\alpha}~~~,
\end{equation}
 together with the adjoint convention $(\chi_1^{\dagger}\chi_2)^{\dagger}=\chi_2^{\dagger}\chi_1$ for Grassmann variables
 $\chi_1,\,\chi_2$, it is easy to verify that $S$ is self-adjoint.

From here on we will usually not indicate the spinor indices $\alpha,\, \beta$ and internal symmetry indices $u,\,v$ explicitly, but they are implicit in all
formulas.   Varying $S$ with respect to the
Rarita-Schwinger fields, we get the equations of motion
\begin{align}\label{eq:eqmo}
\epsilon^{\mu\eta\nu\rho} \partial_{\nu}\overline{\psi}_{\rho}\gamma_{\eta}=& g\epsilon^{\mu\eta\nu\rho}
\overline{\psi}_{\rho}A_{\nu}^A t_{A} \gamma_{\eta}~~~,\cr
\epsilon^{\mu\eta\nu\rho}\gamma_{\eta}\partial_{\nu}\psi_{\rho}=&-g \epsilon^{\mu\eta\nu\rho}
\gamma_{\eta} A_{\nu}^A t_{A} \psi_{\rho}~~~.\cr
\end{align}
Re-expressed in terms of the covariant derivative, these are
\begin{align}\label{eq:eqmo1a}
\epsilon^{\mu\eta\nu\rho} \overline{\psi}_{\rho}\overleftarrow{D}_{\nu}\gamma_{\eta}=&0~~~,\cr
\epsilon^{\mu\eta\nu\rho}\gamma_{\eta}D_{\nu}\psi_{\rho}=&0~~~.\cr
\end{align}

The $\mu=0$ component of these equations gives the primary constraints
\begin{align}\label{eq:constraint1}
\epsilon^{enr} \overline{\psi}_{r}\overleftarrow{D}_{n}\gamma_{e}=&0~~~,\cr
\epsilon^{enr} \gamma_{e}D_{n}\psi_{r}=&0~~~,\cr
\end{align}
with $e,n,r$ summed from 1 to 3.
Contracting the equation of motion for $\overline{\psi}_{\rho}$  with $g^{-1}\overleftarrow{D}_{\mu}$ and
the equation of motion for $\psi_{\rho}$ with $g^{-1}D_{\mu}$, we get the secondary constraints
\begin{align}\label{eq:constraint2}
\epsilon^{\mu\eta\nu\rho} \overline{\psi}_{\rho}F_{\mu\nu}\gamma_{\eta}=&0~~~,\cr
\epsilon^{\mu\eta\nu\rho}\gamma_{\eta}F_{\mu\nu}\psi_{\rho}=&0~~~,\cr
\end{align}
 where we have introduced the  gauge field strength
\begin{align}\label{eq:gaugefield}
F_{\mu\nu}=&g^{-1}[D_{\mu},D_{\nu}]=g^{-1}[\overleftarrow{D}_{\mu},\overleftarrow{D}_{\nu}]\cr
=&\partial_{\mu}A_{\nu}-\partial_{\nu}A_{\mu}+g[A_{\mu},A_{\nu}]~~~,\cr
\end{align}
which with the adjoint representation index $A$ indicated explicitly reads
\begin{equation}\label{eq:gaugefield1}
F_{\mu\nu}^A=\partial_{\mu}A_{\nu}^A-\partial_{\nu}A_{\mu}^A+g f_{ABC}A_{\mu}^B A_{\nu}^C~~~.
\end{equation}
Under a Rarita-Schwinger gauge transformation (with $\epsilon$ a four-component spinor), which is
a natural gauge field generalization of the fermionic gauge invariance for a free, massless Rarita-Schwinger field discussed
in \cite{dasb},
\begin{align}\label{eq:gaugetrans}
\psi_{\mu} \to&  \psi_{\mu}+ \delta_G \psi_{\mu}~~,~~~   \delta_G \psi_{\mu}   \equiv      D_{\mu}\epsilon~~~,\cr
\overline{\psi}_{\mu} \to &  \overline{\psi}_{\mu}+ \delta_G  \overline{\psi}_{\mu} ~~,~~~ \delta_G  \overline{\psi}_{\mu} \equiv  \overline{\epsilon} \overleftarrow{D}_{\mu}~~~,\cr
\end{align}
the action of Eq. \eqref{eq:action} changes according to
\begin{equation}\label{eq:actionchange}
\delta_G S(\psi_{\mu})= -\frac{1}{4}ig \int d^4x \Big[ \overline{\epsilon} \gamma_5 \Big(\epsilon^{\mu\eta\nu\rho} \gamma_{\eta} F_{\mu\nu} \psi_{\rho}\Big)
+\Big(\epsilon^{\mu\eta\nu\rho} \overline{\psi}_{\rho} F_{\mu\nu} \gamma_{\eta}\Big) \gamma_5 \epsilon\Big]+O(\overline{\epsilon}...\epsilon) ~~~.
\end{equation}
The factors bracketed in large parentheses are identical to the secondary constraints of Eq. \eqref{eq:constraint2}. This equation holds with
finite (not necessarily infinitesimal) $\epsilon$ and its adjoint $\epsilon^{\dagger}$; the precise form of the quadratic term is given in Eq. \eqref{eq:fourcompaction} below.
  We will argue in Sec. 5
 that Eq. \eqref{eq:actionchange}  implies that, even when coupled to gauge fields,  the Rarita-Schwinger theory has a generalized form of fermionic gauge invariance.

Adding the gauge field action
\begin{equation}\label{eq:gaugeaction}
S(A_{\mu}^A)=-\frac{1}{4}\int d^4x F_{\mu\nu}^A F^{A\mu\nu}~~~,
\end{equation}
and varying the sum $S(\psi_{\mu})+S(A_{\mu}^A)$ with respect to the gauge potential, we get the
gauge field equation of motion
\begin{align}\label{eq:gaugemo}
D_{\nu}F^{A\mu\nu}&\equiv\partial_{\nu}F^{A\mu\nu}+g f_{ABC} A_{\nu}^B F^{C\mu\nu}=gJ^{A\mu}~~~,\cr
J^{A\mu}=&\frac{1}{2}\overline{\psi}_{\nu}i\epsilon^{\nu\eta\mu\rho}\gamma_5\gamma_{\eta}t_{A} \psi_{\rho}~~~.\cr
\end{align}
A straightforward calculation using Eqs. \eqref{eq:eqmo} shows that the gauge field source current $J^{A\mu}$ obeys the
covariant conservation equation
\begin{equation}\label{eq:jcons}
D_{\mu}J^{A\mu}=\partial_{\mu}J^{A\mu}+gf_{ABC}A_{\mu}^B J^{C \mu}=0~~~,
\end{equation}
as required for consistency of Eq. \eqref{eq:gaugemo}.
So from the Rarita-Schwinger and gauge field actions, we have obtained a formally consistent set of equations of motion.

In addition to the gauge field source current, there is an additional current $J^{\mu}$ that obeys an ordinary conservation
equation,
\begin{align}\label{eq:fermion}
J^{\mu}=&\frac{1}{2}\overline{\psi}_{\nu}\epsilon^{\nu\eta\mu\rho}\gamma_5\gamma_{\eta}\psi_{\rho}~~~,\cr
\partial_{\mu}J^{\mu}=&0~~~.\cr
\end{align}
In the massive spinor case, Velo and Zwanziger \cite{velo} argue that the analogous current, within the constraint subspace of Eq. \eqref{eq:constraint1}, should have a positive time component. In  the massless case we see no reason for this requirement, since
Eq. \eqref{eq:fermion} is the fermion number current, and its time component, giving the fermion number density, can have either sign. However, we shall use parts of the positivity
argument of \cite{velo} later on in discussing positivity of the Dirac bracket anticommutator.

The symmetric stress-energy tensor for the free massless Rarita-Schwinger has been computed by Das \cite{das1} (see also Allcock and Hall \cite{allcock}).   Changing ordinary derivatives to
gauge covariant derivatives, Das's formula becomes
\begin{align}\label{eq:rstensor}
T_{\rm RS}^{\sigma\tau}=&-\frac{i}{4}\epsilon^{\lambda\mu\nu\rho}\Big[\overline{\psi}_{\lambda} \gamma_5 (\gamma^{\tau}\delta^{\sigma}_{\mu}+ \gamma^{\sigma} \delta^{\tau}_{\mu}) D_{\nu}\psi_{\rho}\cr
+&\frac{1}{4}\partial_{\alpha}\Big(\overline{\psi}_{\lambda} \gamma_5 \gamma_{\mu}([\gamma^{\alpha},\gamma^{\sigma}]\delta^{\tau}_{\nu}
+ [\gamma^{\alpha},\gamma^{\tau}]\delta^{\sigma}_{\nu})\psi_{\rho}\Big)\Big]~~~.\cr
\end{align}
\big(This formula can be made manifestly self-adjoint by replacing $D_{\nu}$ by $\frac{1}{2}(D_{\nu}-\overleftarrow{D}_{\nu})$, but this is not needed to verify
stress-energy tensor conservation.\big)  Adding the gauge field stress-energy tensor,
\begin{equation}\label{eq:gauge}
T_{\rm gauge}^{\sigma\tau}=-\frac{1}{4}\eta^{\sigma\tau}F^A_{\lambda\mu}F^{A\lambda\mu} + F^{A\sigma}_{\lambda}F^{A\lambda\tau}~~~,
\end{equation}
a lengthy calculation, using Eq. \eqref{eq:jcons} together with identities and alternative forms of the equations of motion given in Appendix A, shows that the total tensor is conserved,
\begin{equation}\label{eq:conservation}
\partial_{\sigma} (T_{\rm RS}^{\sigma\tau}+T_{\rm gauge}^{\sigma\tau})=0~~~.
\end{equation}

\subsection{Generalization to general $g_{\mu\nu}$}

The generalization of the Rarita-Schwinger action to curved spacetime has been reviewed by Deser and Waldron \cite{deser}.  In Eq. \eqref{eq:action},
$d^4 x$ is replaced by the invariant volume element $d^4x (-g)^{1/2}$, and the covariant derivative $D_{\nu}$ becomes the
curved spacetime covariant derivative
\begin{equation}\label{eq:curvedderiv}
D_{\nu}\psi_{\rho}=\partial_{\nu}\psi_{\rho}-\Gamma_{\nu\rho}^{\beta}\psi_{\beta}+\frac{1}{4}\omega_{\nu m n}\gamma^{m n} \psi_{\rho}
+ gA_{\nu}\psi_{\rho}~~~,
\end{equation}
with $\Gamma_{\nu\rho}^{\beta}$ and $\omega_{\nu m n}$ the affine and spin connections.  The Rarita-Schwinger equation of Eq. \eqref{eq:eqmo1a} and
the primary constraint of Eq. \eqref{eq:constraint1} have the same form as before, in terms of the extended covariant derivative $D_{\nu}$.
The secondary constraint of Eq. \eqref{eq:constraint2} now reads
\begin{align}\label{eq:gravconstraint2}
\epsilon^{\mu\eta\nu\rho} \overline{\psi}_{\rho}[\overleftarrow{D}_{\mu},\overleftarrow{D}_{\nu}]\gamma_{\eta}=&0~~~,\cr
\epsilon^{\mu\eta\nu\rho}\gamma_{\eta}[D_{\mu},D_{\nu}]\psi_{\rho}=&0~~~,\cr
\end{align}
with $\overleftarrow{D}_{\nu}$ defined by  the adjoint of $D_{\nu}$.
The commutator of covariant derivatives is now given by \cite{deser}
\begin{equation}\label{eq:covderivcomm}
[D_{\mu},D_{\nu}]\psi_{\rho}=-R_{\mu\nu\rho}^{\sigma}\psi_{\sigma}+\frac{1}{4}R_{\mu\nu m n} \gamma^{mn}\psi_{\rho}+ g F_{\mu \nu} \psi_{\rho}~~~,
\end{equation}
with $R_{\mu\nu\rho}^{\sigma}$ and $R_{\mu\nu m n}$ components of the Riemann curvature tensor, and as in flat spacetime involves only $\psi_{\rho}$ and not its time or space derivatives.  In terms of the extended covariant derivative, the fermionic gauge transformation is still given by Eq.
\eqref{eq:gaugetrans}, and under this gauge transformation the change in the action is now given by
\begin{equation}\label{eq:gravactionchange}
\delta_G S(\psi_{\mu})= -\frac{1}{4}i \int d^4x \Big[ \overline{\epsilon} \gamma_5 \Big(\epsilon^{\mu\eta\nu\rho} \gamma_{\eta} [D_{\mu},D_{\nu}]\psi_{\rho}\Big)
+\Big(\epsilon^{\mu\eta\nu\rho} \overline{\psi}_{\rho} [\overleftarrow{D}_{\mu},\overleftarrow{D}_{\nu}]\gamma_{\eta}\Big) \gamma_5 \epsilon\Big]+O(\overline{\epsilon}...\epsilon) ~~~,
\end{equation}
with the factors bracketed in large parentheses now identical to the secondary constraints of Eq. \eqref{eq:gravconstraint2} (and again with
  $\epsilon$ and $\epsilon^{\dagger}$ finite).  The arguments
 to be given in Sec. 5 then imply that in the presence of both gravitation and gauge fields,
the Rarita-Schwinger theory has a generalized form of fermionic gauge invariance.  Having established this curved spacetime generalization, we
will continue in the remainder of this and the following paper to work in flat spacetime, but we expect everything done in what follows to
have a curved spacetime generalization when the covariant derivative is suitably extended.

\section{Lagrangian analysis for left chiral spinors in two-component form }

Although we could continue with the four-component formalism to study constraints, the Hamiltonian formalism, and quantization,
it will be more convenient to first reduce the four component equation to decoupled equations for left and right chiral
components of $\psi_{\mu}^{\alpha}$ (with $\alpha$ the spinor index and with the internal symmetry index implicit).  Since these are related by symmetry, we can then focus our analysis on the two-component
equations for the left chiral component, which is the component conventionally used in formulating grand unified
models (see, e.g. \cite{adler}).

We  convert the action of Eq. \eqref{eq:action} to two-component form for the left chiral components of $\psi_{\mu}^{\alpha}$, using the Dirac matrices given in Eqs. \eqref{a2} and \eqref{a4}. Defining the two-component four vector spinor $\Psi_{\mu}^{\alpha}$ and its adjoint $\Psi_{\mu \alpha}^{\dagger}$ by
\begin{align}\label{eq:Psidef}
 P_L \psi_\mu^{\alpha}=&\left( \begin{array} {c}
 \Psi_\mu^{\alpha}  \\
 0 \\  \end{array}\right)~,~~ \mu=0,1,2,3  ~,~~ \alpha=1,2~~~,\cr
 \psi_{\mu\alpha}^{\dagger} P_L=&\Big(\Psi_{\mu \alpha}^{\dagger}~~~0\Big)~~~,\cr
\end{align}
the  action decomposes into uncoupled left and right chiral parts. The left chiral part, with spinor indices $\alpha$ suppressed,  is given by
\begin{equation}\label{eq:leftaction}
S(\Psi_{\mu})=\frac{1}{2}\int d^4x  [-\Psi_{0}^{\dagger} \vec \sigma \cdot \vec D \times \vec {\Psi}
+\vec {\Psi}^{\dagger} \cdot \vec \sigma \times \vec D \Psi_{0}
+\vec{\Psi} ^{\dagger} \cdot \vec D \times \vec \Psi - \vec{\Psi}^{\dagger} \cdot \vec \sigma \times D_{0} \vec{\Psi}]~~~.
\end{equation}
Varying with respect to $\vec{\Psi}^{\dagger}$ we get the Euler-Lagrange equation
\begin{equation}\label{eq:euler}
0=\vec V \equiv \vec \sigma \times \vec D \Psi_{0} + \vec D \times \vec{\Psi}-\vec \sigma \times D_{0}\vec {\Psi}~~~,
\end{equation}
while varying with respect to $\Psi_{0}^{\dagger}$ we get the primary constraint \big(given in four-component form in Eq. \eqref{eq:constraint1}\big)
\begin{equation}\label{eq:chi}
0= V_0\equiv \chi\equiv \vec \sigma \cdot \vec D \times \vec{\Psi}~~~.
\end{equation}
(The abbreviation $V_0\equiv\chi$ conforms to the notation of \cite{velo}.)
A second primary constraint follows from the fact that the action has no dependence on $d\Psi_0^{\dagger}/dt$, which implies that
the  momentum conjugate to $\Psi_0^{\dagger}$ vanishes identically,
\begin{equation}\label{eq:p0}
P_{\Psi_0^{\dagger}}=0~~~.
\end{equation}

Contracting $\vec V$ with $\vec \sigma$ and with $g^{-1}\vec D$, and using the covariant derivative relations of Eq. \eqref{a14}, we
get respectively
\begin{align}\label{eq:omegatheta1}
\vec \sigma \cdot \vec V= &2i \theta +\chi~~~,\cr
g^{-1}\vec D \cdot \vec V=&i \omega+g^{-1} D_0 \chi~~~,\cr
\end{align}
with
\begin{align}\label{eq:omegatheta2}
\theta\equiv&\vec \sigma \cdot \vec D \Psi_0 - D_0 \vec \sigma \cdot \vec \Psi~~~,\cr
\omega\equiv& \vec \sigma \cdot \vec B \Psi_0 -(\vec B + \vec \sigma \times \vec E) \cdot \vec \Psi~~~.\cr
\end{align}
Since the Euler-Lagrange equations imply that $\vec V$ and $\chi$ vanish for all times, we learn that $\theta$ and $\omega$
vanish also for all times.  Since $\theta$ involves a time derivative, its vanishing is just one component of the equation
of motion for $\Psi_{\mu}$.  But $\omega$ involves no time derivatives, so it is a secondary constraint that relates $\Psi_0$  to
$\vec \Psi$ \big(given in four-component form in Eq. \eqref{eq:constraint2}\big).   For each of the above equations, there is a corresponding relation for the adjoint quantity.

The equation of motion $\vec V=0$ can be written in a simpler form by using the identities of Eqs. \eqref{a10} and \eqref{a11} as
follows.  Using Eq. \eqref{a10} to simplify $0=\vec \sigma \times \vec V- i \vec V$, we get an equation for $D_0 \vec \Psi$,
\begin{equation}\label{eq:d0psi1}
D_0\vec \Psi= \vec D \Psi_0+ \frac{1}{2} [-\vec \sigma \times (\vec D \times \vec{\Psi}) + i \vec D \times \vec \Psi]~~~.
\end{equation}
A further simplification can be achieved by incorporating the primary constraint $\chi=0$, through applying Eq. \eqref{a11} to
$\vec A=\vec D \times \vec \Psi$,
\begin{equation}\label{eq:simpli}
0=\vec \sigma ~\chi = \vec \sigma ~\vec \sigma \cdot(\vec D \times \vec \Psi)=\vec D \times \vec \Psi - i \vec \sigma \times (\vec D \times \vec \Psi)~~~.
\end{equation}
Using this to replace the first term in square brackets in Eq. \eqref{eq:d0psi1} we get the alternative
form of the equation of motion, valid when the constraint $\chi=0$ is satisfied,
\begin{equation}\label{eq:d0psi2}
D_0 \vec \Psi=\vec D \Psi_0+ i \vec D \times \vec \Psi~~~.
\end{equation}

Writing the gauge field interaction terms in Eq. \eqref{eq:leftaction} in the form
\begin{equation}\label{eq:interaction}
S_{\rm int} (\Psi_{\mu})= \frac{g}{2} \int d^4 x (A_0^B J^{B0}+\vec A^B \cdot {\vec J}^B)~~~,
\end{equation}
we find the left chiral contribution to the current of Eq. \eqref{eq:gaugemo} in the form
\begin{align}\label{eq:currents}
J^{A0}=&-\vec{\Psi}^{\dagger} t_A \cdot \vec \sigma \times \vec{\Psi}~~~,\cr
\vec{J}^A=& \Psi_0^{\dagger} t_A \vec \sigma \times \vec \Psi+ \vec {\Psi}^{\dagger} \times \vec \sigma t_A \Psi_0 -
\vec{\Psi}^{\dagger} \times t_A \vec{\Psi}~~~.\cr
\end{align}
Replacing $t_A$ by $-i$, we find the corresponding singlet current in the form
\begin{align}\label{eq:currents1}
J^{0}=&i\vec{\Psi}^{\dagger}  \cdot \vec \sigma \times \vec{\Psi}~~~,\cr
\vec{J}=&-i (\Psi_0^{\dagger}  \vec \sigma \times \vec \Psi+ \vec {\Psi}^{\dagger} \times \vec \sigma \Psi_0 -
\vec{\Psi}^{\dagger} \times \vec{\Psi})~~~.\cr
\end{align}
For the energy integral computed from the left chiral part of the the stress-energy tensor of Eq. \eqref{eq:rstensor}, we find
\begin{equation}\label{eq:graven}
H=-\int d^3x T_{RS}^{00}= -\frac{1}{2} \int d^3x \vec{\Psi}^{\dagger} \cdot \vec{D} \times \vec{\Psi}~~~.
\end{equation}

To conclude this section, we verify that the action of Eq. \eqref{eq:leftaction} has a fermionic gauge invariance on the constraint surface
$\omega=0~,~~ \omega^{\dagger}=0$, as already seen in covariant form following Eq. \eqref{eq:gaugetrans}.   Letting $\epsilon$ be a general space and time dependent two-component spinor, we introduce the fermionic gauge changes
\begin{align}\label{eq:gaugedef}
\vec{\Psi} \to & \vec {\Psi} + \delta_G \vec \Psi ~~,~~\delta_G \vec \Psi\equiv \vec D \epsilon~~~,\cr
\Psi_0 \to & \Psi_0 + \delta_G \Psi_0~~,~~  \delta_G \Psi_0 \equiv  D_0 \epsilon~~~, \cr
\end{align}
and their adjoints,
which are the left chiral form of the gauge change of Eq. \eqref{eq:gaugetrans}.   Substituting this into Eq. \eqref{eq:leftaction}, integrating by parts where needed,
and using Eqs. \eqref{a14} to simplify commutators of
covariant derivatives, we find that Eq. \eqref{eq:actionchange} takes the two-component spinor form
\begin{equation}\label{eq:actionchange1}
\delta_G S(\Psi_{\mu}) = \frac{1}{2}ig \int d^4 x (\omega^{\dagger}\epsilon -\epsilon^{\dagger} \omega)+O(\epsilon^{\dagger}...\epsilon)~~~,
\end{equation}
with the quadratic term given in Eq. \eqref{eq:generalaction} below.
Hence the  action on the constraint surface $\omega=\omega^{\dagger}=0$ has a fermionic gauge invariance.
Another gauge invariant, on the constraint surface $\chi=\chi^{\dagger}=0$,  is the fermion number, given
by the space integral of the time component of the singlet current of Eq. \eqref{eq:currents1}, $\int d^3x J^0 $,
which has the gauge variation
\begin{equation}\label{eq:fermionnumbervar}
\delta_G\int d^3x J^0=\int d^3x[-i (\epsilon^{\dagger}\chi+\chi^{\dagger}\epsilon)+g \epsilon^{\dagger} \vec \sigma \cdot
\vec B \epsilon]~~~.
\end{equation}
Again, these equations hold for $\epsilon$ and it adjoint $\epsilon^{\dagger}$ finite.

However, neither the equation of motion, the constraints $\chi$ and $\omega$, the non-Abelian ``charge''  $\int d^3x J^{B0}$,  nor the integrated Hamiltonian $H$ are gauge invariant
in the interacting case. Using $\delta_G$ to denote gauge variations, we have
\begin{align}\label{eq:nongaugeinv}
\delta_G \vec V=&-ig(\vec B + \vec \sigma \times \vec E) \epsilon~~~,\cr
\delta_G \theta =& -ig \vec \sigma \cdot \vec E \epsilon~~~,\cr
\delta_G \chi=&-ig \vec \sigma \cdot \vec B \epsilon~~~,\cr
\delta_G \omega=& \vec \sigma \cdot \vec B D_0 \epsilon - (\vec B+ \vec \sigma \times \vec E) \cdot \vec D \epsilon~~~,\cr
\delta_G \int d^3x J^{B0}=&g\int d^3x \Big( \epsilon^{\dagger} [\vec A, t_B] \cdot \vec \sigma \times \vec \Psi
+ \vec \Psi^{\dagger} \times \vec \sigma \cdot [t_B,\vec A] \epsilon  \Big)~~~,\cr
\delta_G H=& \frac{1}{2}ig \int d^3x (\vec \Psi^{\dagger} \cdot \vec B \epsilon - \epsilon^{\dagger} \vec B \cdot \vec \Psi) ~~~.\cr
\end{align}
The {\it only} global fermionic gauge invariants are the action integral, and the fermion number integral, in both flat and curved
spacetimes.

These results have an interpretation in terms of the distinction between a gauge transformation, customarily defined as an invariance
of the physical state of the system, and a canonical transformation.  The usual  gauge transformations in gauge field theories and
general relativity are invariances of the action without imposition of a constraint, and consequently are invariances of the field
equations and the Hamiltonian.  Such gauge transformations are a special case of canonical transformations, but the converse is not
true:  canonical transformations in general alter the action, the field equations, and the Hamiltonian.  We will see in Sec. 4 that
the fermionic gauge transformation of Eq. \eqref{eq:gaugedef} are always canonical transformations, which reduce to gauge transformations
of the customary type only when the external gauge fields vanish.  However, by virtue of the Jacobi identity for the Poisson
bracket, canonical transformations preserve inner properties of the theory.  As an example, that will be needed in our further
discussion of generalized fermionic gauge invariance in Sec. 5, we verify that the secondary constraint following from the
gauge-varied equation of motion $\vec V$ and primary constraint $V_0=\chi$ agrees with the gauge variation of the
original secondary constraint $\omega$. From Eq. \eqref{eq:omegatheta1} we have
\begin{equation}\label{eq:newomega}
\vec D \cdot \vec V-D_0 \chi=ig\omega~~~.
\end{equation}
Preservation of inner properties under the fermionic gauge transformation means that we should find that
\begin{equation}\label{eq:newomega1}
\vec D \cdot\delta_G  \vec V-D_0\delta_G  \chi=ig\delta_G \omega~~~.
\end{equation}
Substituting Eqs. \eqref{eq:nongaugeinv} into the left hand side of Eq. \eqref{eq:newomega1} gives
\begin{equation}\label{eq:inner1}
ig[D_0 \vec \sigma \cdot \vec B \epsilon-\vec D \cdot (\vec B + \vec \sigma \times \vec E) \epsilon)]=
ig[\vec \sigma \cdot \vec B D_0 \epsilon - (\vec B+ \vec \sigma \times \vec E) \cdot \vec D \epsilon+C \epsilon]~~~,
\end{equation}
with the commutator remainder $C$ given by
\begin{equation}\label{eq:inner2}
C=\vec \sigma \cdot [D_0 \vec B -\vec B D_0 + \vec D \times \vec E +\vec E \times \vec D]
-(\vec D \cdot \vec B - \vec B \cdot \vec D)=0~~~,
\end{equation}
which vanishes by virtue of the gauge field Bianchi identity.

In Sec. 5 we will discuss in more detail why  the fermionic gauge transformation, because it leaves the
constrained action invariant, corresponds to an unwanted redundancy in the time evolution.
To break the gauge invariance we can introduce an additional constraint, in the form
\begin{equation}\label{eq:constraint}
f(\vec \Psi)=0~~~,
\end{equation}
with $f$ a scalar function of its argument.  This constraint, together with the $\chi$ constraint, leaves one independent
two-component spinor of the original three in $\vec \Psi$, corresponding to the physical massless Rarita-Schwinger
modes propagating in the gauge field background.  We will limit ourselves to considering linear constraints of the general form
\begin{equation}\label{eq:constraintform}
f=\vec L \cdot \vec \Psi~~~,
\end{equation}
and the choice $\vec L=\vec D$, a gauge covariant radiation gauge analog,  plays a special role in our analysis.  By not specializing $\vec L$
in our formulas, we can also examine the consequences of omitting a gauge fixing condition, corresponding to taking
$\vec L=0$.

We proceed to examine the gauge covariant radiation gauge condition in more detail.
We note that since
\begin{equation}\label{eq:chisimp}
\vec \sigma \cdot \vec D \vec \sigma \cdot \vec{\Psi}=\vec D \cdot \vec{\Psi}+ i\chi,
\end{equation}
the primary constraint $\chi=0$ implies that
\begin{equation}\label{eq:chisimp1}
\vec \sigma \cdot \vec D \vec \sigma \cdot \vec{\Psi}=\vec D \cdot \vec{\Psi}~~~.
\end{equation}
Hence when $\vec \sigma \cdot \vec D$ is invertible, which is expected in a perturbation expansion in the
gauge coupling $g$, the covariant radiation gauge constraint $\vec D \cdot \vec \Psi=0$ implies that
\begin{equation}\label{eq:sigmaconstraint}
\vec \sigma \cdot \vec \Psi=0~~~.
\end{equation}
Conversely, Eqs. \eqref{eq:chisimp} and \eqref{eq:chisimp1} show that $\vec D \cdot \vec \Psi=0$  and $\vec \sigma \cdot \vec \Psi=0$
together imply the primary constraint $\chi=0$, and also $\vec \sigma \cdot \vec \Psi=0$ and $\chi=0$ imply $\vec D \cdot \vec \Psi=0$.

We next note that on a given initial time slice, covariant radiation gauge is attainable.  Under the gauge
transformation of Eq. \eqref{eq:gaugedef}, we see that
\begin{equation}\label{eq:radgaugechange}
\vec D \cdot \vec \Psi \to \vec D \cdot \vec \Psi+ (\vec D)^2  \epsilon~~~.
\end{equation}
Hence when $(\vec D)^2$ is invertible, which we expect to be true in a perturbative
sense, then we can invert $(\vec D)^2 \epsilon = - \vec D \cdot \vec \Psi$, to find
a gauge function $\epsilon$ that brings a general $\vec \Psi$ to covariant radiation
gauge.  Since
\begin{equation}\label{eq:sigmadsquared}
(\vec \sigma \cdot \vec D)^2=(\vec D)^2+g \vec \sigma \cdot \vec B~~~,
\end{equation}
the conditions for $\vec \sigma \cdot \vec D$ to be invertible, and
for $(\vec D)^2$ to be invertible, are  related.  For generic non-Abelian
gauge fields both of these operators should be invertible, but there will be
isolated gauge field configurations for which $\vec \sigma \cdot \vec D$ has zeros.

However, although covariant radiation gauge can be imposed on any time slice, it is
not preserved by the equation of motion for $\vec \Psi$.  To see this, let us consider
the simplified case in which the gauge potential is specialized to $A_0=0$ and $\partial_0 \vec A=0$,
so that only a static $\vec B$ field is present.  From Eq. \eqref{eq:d0psi2} we have
\begin{equation}\label{eq:d0radgauge}
\partial_0(\vec D \cdot  \vec \Psi)=(\vec D)^2 \Psi_0+ g \vec B \cdot \vec \Psi
 =[(\vec D)^2+g \vec \sigma \cdot \vec B]\Psi_0 = (\vec \sigma \cdot \vec D)^2 \Psi_0~~~.
\end{equation}
So $\partial_0(\vec D \cdot  \vec \Psi)=0$ implies $\Psi_0=0$, but this is one constraint too many.
Hence at each infinitesimal time step, we must make a further infinitesimal fermionic gauge transformation
to maintain the covariant radiation gauge condition, as further discussed in Sec. 5B below.   Only in the absence of gauge fields can we
simultaneously impose the constraints  $\vec \nabla \cdot \vec{\psi}=0$, $\vec \sigma \cdot \vec{\psi}=0$, and
$\psi_0=0$, as used in the discussion of \cite{dasb} for the free Rarita-Schwinger case,

\section{Canonical momenta, classical brackets, and gauge generators}

We next introduce the canonical momentum conjugate to $\vec \Psi$,
defined by
\begin{equation}\label{eq:canmom}
\vec P = \frac{\partial^L S}{\partial (\partial_0 \vec{\Psi}) } = \frac{1}{2} \vec{\Psi}^{\dagger} \times \vec \sigma~~~,
\end{equation}
which can be solved for $\vec{\Psi}^{\dagger}$ using the final line of Eq. \eqref{a11},
\begin{equation}\label{eq:psidagger}
\vec{\Psi}^{\dagger}=i \vec P - \vec P \times \vec \sigma~~~.
\end{equation}
We will use Eq. \eqref{eq:psidagger} when computing classical brackets involving $\vec{\Psi}^{\dagger}$ using the formula of
Eq. \eqref{a17}.
Eq. \eqref{eq:canmom} can be written as an explicit matrix relation for the  six components of $\vec P$ and $\vec{\Psi}^{\dagger}$,
\begin{equation}\label{eq:matrixform}
\left( \begin{array} {c} P_1^{\uparrow}\\P_1^{\downarrow}\\
P_2^{\uparrow} \\P_2^{\downarrow}\\P_3^{\uparrow}\\ P_3^{\downarrow}\\
\end{array}\right)=\frac{1}{2}\left( \begin{array} {cccccc}
 0  & 0 & 1 &0   &0   & -i   \\
 0  & 0 &0  & -1  & i  & 0   \\
-1   &0  &0  &0   &0   &1    \\
 0  &1  &0  &0   &1   & 0   \\
 0  &i  &0  & -1  &0   & 0   \\
 -i  &0 &-1  &0   & 0  & 0   \\
\end{array} \right)
\left( \begin{array} {c} \Psi_1^{\dagger\uparrow}\\ \Psi_1^{\dagger\downarrow}\\
\Psi_2^{\dagger\uparrow} \\ \Psi_2^{\dagger\downarrow}\\ \Psi_3^{\dagger\uparrow}\\ \Psi_3^{\dagger\downarrow}\\
\end{array}\right)~~~,
\end{equation}
showing that they are related by an anti-self-adjoint matrix with determinant $-1/16$.

The four constraints introduced in Sec. 3 are
\begin{align}\label{eq:phicon}
\phi_1=&P_{\Psi_0^{\dagger}}~~~,\cr
\phi_2=&(\vec \sigma \cdot \vec B)^{-1} \omega= \Psi_0-(\vec \sigma \cdot \vec B)^{-1} (\vec B+\vec \sigma \times \vec E)
\cdot \vec\Psi~~~,\cr
\phi_3=&\chi=\vec \sigma \cdot \vec D \times \vec \Psi~~~,\cr
\phi_4=&\vec L \cdot \vec \Psi~~~.\cr
\end{align}
In writing these we are assuming that $\vec \sigma \cdot \vec B$ is invertible in the non-Abelian case.  We are writing the gauge fixing condition as  a general  linear gauge fixing constraint $\vec L\cdot \vec \Psi$, so
as to keep track of which terms in the final answers arise from gauge fixing, which is not evident if we specialize by replacing $\vec L$ by
$\vec D$ at this stage.  The constraints of Eq. \eqref{eq:phicon}, including the gauge fixing
constraint $\phi_4$, are all first class in the Dirac classification, since they have vanishing mutual classical brackets.  This is
a consequence of the fact that starting with a constraint depending on $\vec \Psi$ but not on $\vec \Psi^{\dagger}$, and taking an arbitrary
number of time derivatives, one still has a constraint depending only on $\vec \Psi$.

To preserve the adjoint properties of the Rarita-Schwinger equation, for each of these four constraints we must impose
 a corresponding adjoint constraint.  Using Eq. \eqref{eq:psidagger} to express $\vec{\Psi}^{\dagger}$
in terms of $\vec P$, we write these as
\begin{align}\label{eq:chidefs}
\chi_1=&(P_{\Psi_0^{\dagger}})^{\dagger}=-P_{\Psi_0}~~~,\cr
\chi_2=&\omega^{\dagger}(\vec \sigma \cdot \vec B)^{-1}=\Psi_0^{\dagger}- \vec P \cdot [i(\vec B+\vec \sigma \times \vec E)
-  \vec \sigma \times (\vec B+\vec \sigma \times \vec E)](\vec \sigma \cdot \vec B)^{-1}~~~,\cr
\chi_3=&\chi^{\dagger}=2 \vec P \cdot \overleftarrow D ~~~,\cr
\chi_4=& \vec{\Psi}^{\dagger} \cdot \overleftarrow{L}= \vec P \cdot (i \overleftarrow{L} -  \vec \sigma \times \overleftarrow{L})~~~.\cr
\end{align}
(The reason for the minus sign in the definition  $P_{\Psi_0^{\dagger}}=-P_{\Psi_0}^{\dagger}$ will be given in Sec. 2 of the following
 paper where we discuss the Hamiltonian
form of the equations.)  The constraints $\phi_a$ are implicitly $2n$ component column vectors, and the adjoint constraints $\chi_a$ are implicitly $2n$ component row vectors,
with $2n$ arising from the product of a factor of 2 for the two implicit spinor indices, and a factor of $n$ for the $n$  implicit $SU(n)$  internal symmetry indices.

When $\vec L=\vec D$, we see that $\phi_4$ becomes $\phi_4=\vec D \cdot \vec \Psi$, and
$\chi_4$ becomes $\chi_4=i\vec P \cdot \overleftarrow{D} - \vec P \cdot \vec \sigma \times \overleftarrow{D} = (i/2) \chi_3 -\vec P \cdot \vec \sigma \times \overleftarrow{D}$.
So a special feature of covariant radiation gauge, which will be exploited later, is that the constraints $\phi_3,\,\phi_4$ are contractions
of $\vec \sigma \times \vec D$ and $\vec D$ with $\vec \Psi$, and the constraints $\chi_3,\,\chi_4$ are contractions of linear combinations
of the duals $\overleftarrow{D}$ and $\vec \sigma \times \overleftarrow{D}$ with $\vec P$.  That is, in covariant radiation gauge the
constraint spaces selected by $\chi_3,\chi_4$ and $\phi_3,\phi_4$ are duals of one another.

We can now compute the classical brackets of the constraints.  We see that the brackets of the $\phi$s and $\chi$s vanish among themselves,
\begin{align}\label{eq:vanishbracks}
[\phi_a,\phi_b]_C=&0~~~,\cr
[\chi_a,\chi_b]_C=&0~~~,\cr
a,b=&1,...,4~~~.\cr
\end{align}
On the other hand, the brackets of the $\phi$s with the $\chi$s give a nontrivial matrix of brackets $M$, which has a nonvanishing determinant,
\begin{align}\label{eq:nonvanishbracks}
M_{ab}(\vec x, \vec y) \equiv &  [\phi_a(\vec x),\chi_b(\vec y)]_C \neq 0~~~,\cr
\det{M} \neq & 0~~~.\cr
\end{align}
Thus, in terms of the Dirac classification, the original first class constraints $\phi_a$ have become second class, not from
adding new constraints that follow from differentiation with respect to time or from imposing gauge fixing conditions, but rather
from adjoining the adjoint set of constraints.  This is a feature of the Rarita-Schwinger constrained fermion system that
has no analog in the familiar constrained boson systems such as gauge fields.

Evaluating the brackets shows that $M$ has the general form
\begin{equation}\label{eq:pstructure}
M=\left( \begin{array} {cccc}
 0&-1&0&0 \\
 1&{\cal U}&{\cal S}&{\cal T} \\
 0&{\cal V}&{\cal A}&{\cal B} \\
 0&{\cal W}&{\cal C}&{\cal D} \\
 \end{array} \right)~~~,
\end{equation}
where in the $SU(n)$ gauge field case, each entry in $M$ is a $2n\times 2n $ matrix (corresponding to the fact that $\phi_a$ is implicitly a $2n$ component
column vector, and $\chi_b$ is implicitly a $2n$ component row vector).  Evaluating
$\det{M}$ by a cofactor expansion with respect to the elements of the two unit matrices $\pm 1$, we see that
the submatrices ${\cal U},\, {\cal S},\, {\cal T},\, {\cal V},\,{\cal W}$ do not contribute, and we have
\begin{align}\label{eq:detp}
\det{M}=&\det{N}~~~\cr
N= & \left(\begin{array} {cc}
  {\cal A}&{\cal B} \\
 {\cal C}&{\cal D} \\
 \end{array} \right)   ~~~.\cr
 \end{align}
So we need to only evaluate the brackets $M_{33}={\cal A}$, \, $M_{34}={\cal B}$,\,
$M_{43}={\cal C}$,\, $M_{44}={\cal D}$, giving
\begin{align}\label{eq:abcdformulas}
{\cal A}=& -2ig \vec \sigma \cdot \vec B(\vec x) \delta^3(\vec x-\vec y)~~~, \cr
{\cal B}=&-2 \vec D_{\vec x}\cdot \vec L_{\vec x} \delta^3(\vec x-\vec y)~~~, \cr
{\cal C}=& 2 \vec{L}_{\vec x}\cdot \vec D_{\vec x} \delta^3(\vec x-\vec y)~~~, \cr
{\cal D}=& \big(i (\vec{L}_{\vec x})^2+\vec \sigma \cdot(\vec L_{\vec x} \times \vec L_{\vec x})\big) \delta^3(\vec x-\vec y)~~~. \cr
\end{align}
When $\vec L=\vec D$, these become
\begin{align}\label{eq:abcdformulas1}
{\cal A}=& -2ig \vec \sigma \cdot \vec B(\vec x) \delta^3(\vec x-\vec y)~~~, \cr
{\cal B}=&-2 (\vec D_{\vec x})^2 \delta^3(\vec x-\vec y)~~~, \cr
{\cal C}=& 2 (\vec D_{\vec x})^2 \delta^3(\vec x-\vec y)~~~, \cr
{\cal D}=& i\big( (\vec{D}_{\vec x})^2-g \vec \sigma \cdot \vec B({\vec x})\big) \delta^3(\vec x-\vec y)~~~. \cr
\end{align}

Reflecting the fact that the $\phi_a$ and $\chi_a$ are adjoints of one another, together with
the fact that the matrix relating  $\vec \Psi^{\dagger} $ to $\vec P$ is anti-self-adjoint \big(see Eq. \eqref{eq:matrixform}\big),
these matrix elements obey the adjoint relations
\begin{equation}\label{adjointrels}
M_{ab}(\vec x, \vec y)^{\dagger}=-M_{ba}(\vec y, \vec x)~~~~.
\end{equation}
Applications of these bracket and determinant calculations will be made in the subsequent paper, where we discuss quantization
by both the Dirac bracket formalism and by the Feynman path integral.

To conclude this section, we note that the constraints $\chi,\, \chi^{\dagger},\, P_{\Psi_0},\,
P_{\Psi_0^{\dagger}}$ play the role of gauge transformation generators. For example,
we have (with common time argument $t$ suppressed)
\begin{align}\label{eq:gaugegenex}
\left[\int d^3x \frac{1}{2}\chi^{\dagger}(\vec x)  \epsilon(\vec x), \vec \Psi(\vec y)\right]_C=&\vec  D_{\vec y}\, \epsilon(\vec y)~~~,\cr
\left[-\int d^3x P_{\Psi_0}(\vec x) D_0 \epsilon(\vec x), \Psi_0(\vec y)\right]_C=& D_{0\vec y}\, \epsilon(\vec y)~~~.\cr
\end{align}
So the fermionic gauge transformation is a canonical transformation.  This is also evident from the fact that
since Eq. \eqref{eq:gaugedef} is just a shift in the fermionic variables $\vec \Psi$ and $\Psi_0$ by the quantities $\vec D \epsilon$
and $D_0\epsilon$, which have no dependence on the fermionic variables, this shift leaves the canonical brackets
$[\vec \Psi_i, \vec P_j]_c$, $[\Psi_0,P_{\Psi_0}]_c$ etc. unchanged.

\section{Generalized gauge invariance of the Rarita-Schwinger action}

We turn now to a justification of our claim that the fermionic gauge transformation introduced in Eqs. \eqref{eq:gaugetrans} and  \eqref{eq:gaugedef} is a generalized form of gauge invariance, which corresponds to redundant degrees of freedom, and which leaves essential attributes of the physics of gauged Rarita-Schwinger fields invariant.
In the most familiar gauge invariant theories, such as Abelian or non-Abelian gauge fields, the Lagrangian density
is invariant under a gauge transformation on the fields. These theories exhibit what one could term ``strong'' gauge invariance.  In a weaker form of gauge invariance, which occurs for the free  Rarita-Schwinger
equation, the Lagrangian density changes by a total derivative under a gauge transformation of the fields, and so only the action
is gauge invariant. Characteristic features of this case have been studied by  Das \cite{das1}.  We argue in this section that there is a still weaker form of gauge
invariance, obeyed by the massless Rarita-Schwinger equation with Abelian or non-Abelian gauging, in which under a gauge transformation
the Lagrangian changes by a total derivative plus terms which vanish when initial value constraints are obeyed.

We divide our argument that the transformation of Eqs. \eqref{eq:gaugetrans} and \eqref{eq:gaugedef} is a generalized
form a gauge invariance into two parts, first considering infinitesimal transformations, and then considering general finite
transformations.

\subsection{Infinitesimal gauge transformations}

In his seminal analysis of constrained systems, Dirac  \cite{dirac} classifies as ``first class'' constraints the maximal set of
constraints that have vanishing mutual Poisson brackets, and notes that ``Each of them thus leads to an arbitrary function of the time
in the general solution of the equations of motion with given initial conditions''.  Elaborating on this, he notes that
``Different solutions of the equations of motion, obtained by different choices of the arbitrary functions of the time with given
initial conditions, should be looked upon as all corresponding to the same physical state of motion, described in various way (sic) by
different choices of some mathematical variables that are not of physical significance (e.g. by different choices of the gauge
in electrodynamics or of the co-ordinate system in a relativistic theory.)''

These remarks suggest that gauge invariance, in its most general form, corresponds to an arbitrariness in the time evolution of a system,
in the sense that the future evolution of the system is not uniquely determined by the initial conditions and the Euler-Lagrange equations
following from the action principle.  Under this generalized definition, the Rarita-Schwinger equation with coupling to gauge fields
has a fermionic gauge invariance.  To see this, we note the Euler-Lagrange equations yield equations of two types.  The first are the
time evolution equations contained in Eq. \eqref{eq:eqmo}, that determine the field variables at a later time $t+\Delta t$ from those initially given at time $t$.
The second are the primary and secondary constraints of Eqs. \eqref{eq:constraint1} and \eqref{eq:constraint2}, which constrain the
initial field values at time $t$.  If we make the gauge transformation of Eq. \eqref{eq:gaugetrans} at time $t$, with infinitesimal gauge parameter
$\epsilon$  (with $\epsilon^{\dagger}$ its adjoint), we have seen that
the action at time $t$ changes, to first order in $\epsilon$,  according to Eq. \eqref{eq:actionchange}.  So assuming that the initial data at time $t$ obeys both the primary and secondary constraints, then when the constraints at time $t$ are applied the change in the action is $O\big((\epsilon)^2\big)$.  After this gauge transformation, we have seen in Eq. \eqref{eq:nongaugeinv} that the Euler-Lagrange equations
$\vec V$, the primary constraint $\chi$ and the secondary constraint $\omega$ are all changed at order $\epsilon$, but because the
gauge transformation is a canonical transformation that preserves inner properties, we have also seen that the altered secondary
constraint is the one implied by the altered $\vec V$ and $\chi$, with an error of at most $(\epsilon)^2$.  Hence after the gauge
transformation, we still have consistent equations of motion and initial conditions, which can serve as a starting point for time evolution.
However, by making the gauge infinitesimal gauge transformation, we have introduced an arbitrariness into the evolved solution.
In order to get a unique time evolution path from the initial data at time $t$ using the action principle, one must impose a gauge
fixing condition, that selects one member out of the equivalence class of equal action field configurations.

In the gauged Rarita-Schwinger theory only the constrained action and  constrained fermion number, in both flat and  curved spacetimes, are invariant to first order under infinitesimal fermionic gauge transformations. This has an important physical significance.  Consider a set of Rarita-Schwinger
fields that, as envisaged in the model of \cite{adler}, are permanently bound into condensates.  The only way to see that these
fields are present is through their gravitational fields, through their gauge field polarizabilities, and possibly also through their
influence on overall fermion number counting.  The constrained action is the functional of the metric and the gauge fields that determines the influence of the Rarita-Schwinger fields on the metric and the gauge fields respectively, so the fact that the constrained action is invariant under infinitesimal fermionic gauge transformations means that the physical effects induced by confined Rarita-Schwinger fields are similarly invariant. (This
statement is not contradicted by the fermionic gauge non-invariance of the energy integral and the gauge field source currents, since these are calculated by varying the unconstrained action, and do not take into account the fact that the constraints that enter into the constrained action are themselves non-trivial functions of the spacetime metric and the gauge fields.)

The fermionic gauge invariance of the constrained action functional of the metric and the gauge fields then
allows us to impose a gauge fixing constraint, making the time evolution determined by the action principle unique. Gauge fixing eliminates the redundancy
of gauge degrees of freedom, and so is a convenience in checking  the correct helicity counting for the Rarita-Schwinger fields, but is not
needed for this purpose.  In the following paper, where we turn to quantization, gauge fixing is needed to get an invertible
constraint matrix in the weak field limit, and when covariant radiation gauge fixing is used one finds manifestly positive semi-definite anticommutation relations for the quantized Rarita-Schwinger fields.

\subsection{Finite gauge transformations: auxiliary fields and the extended action}

Since the transformations of Eqs. \eqref{eq:gaugetrans} and \eqref{eq:gaugedef} are linear, and since the Euler-Lagrange equations and
primary and secondary constraints are linear in the Rarita-Schwinger field, the relations of Eq.  \eqref{eq:nongaugeinv} give
the most general form of the transformed equations of motion and constraints.  Thus, letting $\Lambda$ denote a finite fermionic gauge transformation, the general form of the equations of motion and constraints are

\begin{align}\label{eq:generalform}
0=\vec V(\Lambda) =& \vec \sigma \times \vec D \Psi_0+\vec D \times \vec \Psi - \vec \sigma \times D_0 \vec \Psi
-ig (\vec B+\vec \sigma \times \vec E)\Lambda~~~,\cr
0=\chi(\Lambda)=&\vec \sigma \cdot \vec D \times \vec \Psi - ig \vec \sigma \cdot \vec B \Lambda ~~~,\cr
0=\omega(\Lambda)=&\vec \sigma \cdot \vec B(\Psi_0+D_0 \Lambda) - (\vec B + \vec \sigma \times \vec E) \cdot (\vec \Psi+\vec D \Lambda)~~~.\cr
\end{align}
Under the gauge shifts of Eq. \eqref{eq:gaugedef}, $\Lambda$ is augmented to $\Lambda+\epsilon$, or equivalently,
under the extended gauge transformation that includes a shift of $\Lambda$, 
\begin{equation}\label{eq:shiftinv}
\Psi_0 \to \Psi_0+D_0\epsilon~,~~ \vec \Psi \to \vec \Psi+ \vec D \epsilon~,~~\Lambda \to \Lambda -\epsilon~~~,
\end{equation}
the formulas of Eq. \eqref{eq:generalform} are left invariant.  By using Eq. \eqref{eq:inner2}, one can verify that
\begin{equation}\label{eq:inner3}
\vec D \cdot \vec V(\Lambda)-D_0\chi(\Lambda)=ig \omega(\Lambda)~~~.
\end{equation}
From Eqs. \eqref{eq:generalform} one deduces  alternative forms of the $\vec \Psi$ equation of motion,
subject to the constraint $\chi(\Lambda)=0$,
\begin{align}\label{eq:altform}
D_0 \vec \Psi =&\vec D \Psi_0 +i \vec D \times \vec \Psi + g(\vec B-i\vec E) \Lambda~~~,\cr
0=&\theta(\Lambda)\equiv \vec \sigma \cdot \vec D \Psi_0-D_0 \vec \sigma \cdot \vec \Psi-i g\vec \sigma \cdot \vec E \Lambda~~~.
\end{align}
From the first of these one finds
\begin{equation}\label{eq:maintgauge}
D_0 \vec D \cdot \vec \Psi= (\vec D)^2 \Psi_0 + g(\vec B +i\vec E)\cdot \vec \Psi+ g \vec D \cdot \big((\vec B-i\vec E) \Lambda\big)~~~,
\end{equation}
which gives a condition on the gauge shift $\Lambda$ for the covariant radiation gauge condition $\vec D \cdot \vec \Psi=0$ to
be maintained in time.

We can now write down an action corresponding to the generalized equations of motion and constraints.  It is
\begin{align}\label{eq:generalaction}
S(\Lambda)=&
\frac{1}{2}\int d^4x  [-\Psi_{0}^{\dagger} \vec \sigma \cdot \vec D \times \vec {\Psi}
+\vec {\Psi}^{\dagger} \cdot (\vec \sigma \times \vec D \Psi_{0}
+\vec D \times \vec \Psi -  \vec \sigma \times D_{0} \vec{\Psi})\cr
-&ig\vec \Psi^{\dagger} \cdot (\vec B + \vec \sigma \times \vec E) \cdot \Lambda+ig \Lambda^{\dagger}
(\vec B + \vec \sigma \times \vec E) \cdot  \vec \Psi \cr
+&ig \Psi_0^{\dagger} \vec \sigma \cdot \vec B \Lambda -ig \Lambda^{\dagger} \vec \sigma \cdot \vec B \Psi_0\cr
+&ig \Lambda^{\dagger} (\vec B + \vec \sigma \times \vec E) \cdot\vec D \Lambda - ig \Lambda^{\dagger} \vec \sigma \cdot \vec B D_0 \Lambda]~~~.\cr
\end{align}
One can check that the final line of this action is self-adjoint, by using Eq. \eqref{eq:inner2}, and
one can also verify that this action is exactly invariant under the transformation of Eq. \eqref{eq:shiftinv}, including quadratic terms
in $\epsilon$, without using the constraints following from the equations of motion.  The extended action of Eq. \eqref{eq:generalaction} gives the most general form of the gauged Rarita-Schwinger action, in which $\Lambda$ plays the role of an auxiliary field that restores exact fermionic
gauge invariance.

Varying this action with respect to $\Psi^{\dagger}$ gives the generalized equation of motion $\vec V(\Lambda)=0$, while varying it
with respect to $\Psi_0^{\dagger}$ gives the generalized primary constraint $\chi(\Lambda)=0$.  Since these hold for all times,
Eq. \eqref{eq:inner3} then shows that they imply the generalized secondary constraint $\omega(\Lambda)=0$.  Varying this action
with respect to $\Lambda^{\dagger}$ gives just the secondary constraint $\omega(\Lambda)=0$ as the equation of motion
for $\Lambda$.   This shows that  $\Lambda$ is not an independent dynamical variable but rather is a Lagrange multiplier
for the secondary constraint, which plays the role of a generalized  auxiliary field.   This further supports our argument that  the gauge transformation of Eq. \eqref{eq:gaugedef} corresponds to  a generalized gauge invariance, and that the gauge degrees of freedom are redundant degrees of freedom.

Making the shift $\epsilon=-\Lambda$  reduces $\Lambda$ to zero, so that action of Eq. \eqref{eq:generalaction}
reduces to its first line, which is the original action of Eq. \eqref{eq:leftaction}.  Conversely, this shows that Eq.  \eqref{eq:generalaction}
is just Eq. \eqref{eq:leftaction} with the substitutions $\Psi_0 \to \Psi_0 + D_0 \Lambda$ and $\vec \Psi \to \vec \Psi+ \vec D \Lambda$,
that is
\begin{align}\label{eq:generalaction1}
 S(\Lambda)=&
\frac{1}{2}\int d^4x  [-(\Psi_{0}^{\dagger}+\Lambda^{\dagger} \overleftarrow D_0) \vec \sigma \cdot \vec D \times (\vec {\Psi}+\vec D \Lambda)\cr
+&(\vec {\Psi}^{\dagger}+ \Lambda^{\dagger}\overleftarrow D) \cdot \big(\vec \sigma \times \vec D (\Psi_{0}+D_0 \Lambda)
+ \vec D \times (\vec \Psi+\vec D \Lambda) - \vec \sigma \times D_{0} (\vec{\Psi}+\vec D \Lambda)\big)]~~~,\cr
\end{align}
which makes manifest the invariance of $S(\Lambda)$ under the shift transformation of Eq. \eqref{eq:shiftinv}. The simplicity of this way
of constructing the extended action is a reflection of the fact that the fermionic gauge group is simply an Abelian group under addition
of gauge functions.  If we now define $\Psi_0^{\prime} =\Psi_0+ D_0\Lambda$ and
$\vec \Psi^{\prime}=\vec \Psi+\vec D \Lambda$, and fix the choice of $\Lambda$ by imposing a gauge fixing condition, such
as gauge covariant radiation gauge, then we see that as function of the primed, gauge-fixed variables the generalized action $S(\Lambda)$ takes
the same form as the original action of Eq. \eqref{eq:leftaction} took as a function of the original variables.

The above analysis in terms of two-component, left chiral spinors can also be carried out in the original four-component formalism.  Making
the substitution $\psi_{\mu} \to \psi_{\mu}+ D_{\mu}\Lambda$ in Eq. \eqref{eq:action} gives after some algebra using Eq. \eqref{eq:gaugefield}
the four-component form of the extended action functional of the Rarita-Schwinger field $\psi_{\rho}$ and the auxiliary field $\Lambda$,
\begin{align}\label{eq:fourcompaction}
S(\Lambda)= &\frac{i}{2}\int d^4 x \epsilon^{\mu\eta\nu\rho}[\overline{\psi}_{\mu}\gamma_5\gamma_{\eta}D_{\nu} \psi_{\rho} \cr
+&\frac{g}{2}(-\overline{\Lambda} \gamma_5\gamma_{\eta} F_{\mu\nu}\psi_{\rho}+ \overline{\psi}_{\mu} \gamma_5\gamma_{\eta} F_{\nu\rho}\Lambda
-\overline{\Lambda} \gamma_5\gamma_{\eta} F_{\nu\rho} D_{\mu} \Lambda)]~~~.\cr
\end{align}
which is self adjoint by virtue of the Bianchi identity
\begin{equation}\label{eq:covbianchi}
\epsilon^{\mu\eta\nu\rho}[D_{\mu}, F_{\nu\rho}]=0~~~.
\end{equation}
Varying Eq. \eqref{eq:fourcompaction}  with respect to $\overline{\psi}_{\mu}$ gives the generalized Euler-Lagrange equations (which include the generalized primary constraint)
\begin{equation}\label{eq:geneuler}
\epsilon^{\mu\eta\nu\rho}(D_{\nu}\psi_{\rho} + \frac{g}{2}F_{\nu\rho}\Lambda)=0~~~,
\end{equation}
while applying $g^{-1}D_{\mu}$ to this and using Eq. \eqref{eq:covbianchi} gives the generalized secondary constraint
\begin{equation}\label{eq:genseccons}
\epsilon^{\mu\eta\nu\rho}F_{\mu\nu}(\psi_{\rho}+D_{\rho}\Lambda)=0~~~.
\end{equation}
Varying Eq. \eqref{eq:fourcompaction} with respect to $\overline{\Lambda}$ gives just the generalized secondary constraint of
Eq. \eqref{eq:genseccons}, again showing that $\Lambda$  is a Lagrange multiplier for the secondary constraint which acts as
an auxiliary field, and thus
corresponds to a redundant degree of freedom, not a physical degree of freedom.

\section{Propagation of a Rarita-Schwinger field in an external Abelian gauge field:  absence of superluminal propagation}

We  specialize now to the case of a Rarita-Schwinger spinor propagating in an external Abelian gauge field, as studied by
Velo and Zwanziger \cite{velo}.  For an Abelian gauge field,
\begin{equation}\label{eq:sigbinversion}
\frac{1}{\vec \sigma \cdot \vec B}=\frac{\vec \sigma \cdot \vec B}{(\vec B)^2}~~~,
\end{equation}
and so $\vec \sigma \cdot \vec B$ is invertible as long as $(\vec B)^2>0$, which we assume.   Provided the Lorentz invariant expression $(\vec B)^2-(\vec E)^2$ is positive, $(\vec B)^2$ will be positive in any Lorentz frame.   In discussing undamped wave propagation we will not use the inequality $(\vec B)^2-(\vec E)>0$, but in treating damped longitudinal mode propagation in Appendix B, we
will assume that $(\vec E)^2/(\vec B)^2$ is small, as motivated by the fact that when $(\vec E)^2$ is of order
 $(\vec B)^2$ the vacuum is highly unstable against pair creation. \big(Strictly speaking, the vacuum is stable against
pair production only when $\vec E \cdot \vec B=0$ and $(\vec B)^2 -(\vec E)^2 >0$, that is, when there is
a Lorentz frame in which the Abelian field has vanishing $\vec E$ \cite{schwinger}.\big)

Given that $(\vec B)^2>0$, we can solve the constraint $\omega=0$ of Eq. \eqref{eq:omegatheta2} for
$\Psi_0$, giving
\begin{equation}\label{eq:solve1}
\Psi_0=\frac{\vec Q  \cdot \vec \Psi}{(\vec B)^2}~~~,
\end{equation}
where we have defined
\begin{equation}\label{eq:qdef}
\vec Q\equiv  \vec \sigma \cdot \vec B (\vec B + \vec \sigma \times \vec E)=
\vec B\times \vec E + \vec B \vec \sigma \cdot (\vec B +i \vec E)-i \vec B \cdot \vec E \vec \sigma~~~.
\end{equation}
Substituting the solution for $\Psi_0$ into Eq. \eqref{eq:d0psi2}, we get an equation of motion for $\vec \Psi$
by itself,
\begin{equation}\label{eq:d0psi3}
D_0 \vec \Psi=\vec D \frac{\vec Q  \cdot \vec \Psi}{(\vec B)^2}+ i \vec D \times \vec \Psi~~~.
\end{equation}

To determine the wave propagation velocity in the neighborhood of a spacetime point $x_*=(t_*,\vec x_*)$, we need to calculate the equation for
the wavefronts, or characteristics, at that point.   Writing the first order  Eq. \eqref{eq:d0psi3} in the form
\begin{equation}\label{eq:d0psi4}
\partial_0 \vec \Psi=\vec \nabla \frac{\vec Q_*  \cdot \vec \Psi}{(\vec B_*)^2}+ i \vec  \nabla\times \vec \Psi+
\vec \Delta[\vec \Psi,  x_*, x]~~~,
\end{equation}
with $\vec B_*$ and $\vec Q_*$ the values of the respective quantities at $ x_*$,
we see that $\vec \Delta[\vec \Psi,  x_*, x]$ involves no first derivatives of $\vec \Psi$ at  $ x_*$, and so is not needed \cite{courant},
 \cite{madore} for determining the wavefronts of  Eq. \eqref{eq:d0psi2}.  The reason  is that when taking an infinitesimal line integral
of Eq. \eqref{eq:d0psi4}, according to
\begin{equation}\label{eq:lineint}
\lim_{\delta \to 0} \int_{-\delta}^{\delta} d\ell [\partial_0 \vec \Psi = ...]~~~,
\end{equation}
discontinuities across wavefronts contribute through the first derivative terms, but when the external fields are smooth
the term $\vec \Delta[\vec \Psi,  x_*, x]$  makes a vanishing contribution as $\delta \to 0$.
Dropping $\vec \Delta$, and multiplying through by $(\vec B_*)^2$, we get the equation determining the wavefronts in the form
\begin{equation}\label{eq:d0psifinal}
(\vec B_*)^2 \partial_0 \vec \Psi=\vec \nabla \vec Q_*  \cdot \vec \Psi+ i (\vec B_*)^2 \vec  \nabla\times \vec \Psi~~~.
\end{equation}
By similar reasoning, the constraint $\chi$ can be simplified, for purposes of determining the wavefronts, by replacing
$\vec D$ by $\vec \nabla$, giving
\begin{equation}\label{eq:newchi}
0=\vec \sigma \cdot \vec \nabla   \times \vec{\Psi}~~~.
\end{equation}

Since these are now linear equations with constant coefficients, the solutions are plane waves, and without loss of generality we can take the negative $z=x_3$ axis as the direction of wave propagation.  So making the Ansatz
\begin{equation}\label{eq:ansatz}
\vec \Psi= \vec C \exp(i \Omega t + i K z)~~~,
\end{equation}
 Eq. \eqref{eq:d0psifinal} for the wavefronts or characteristics takes the form
\begin{equation}\label{eq:ceq}
0=\vec F\equiv (\vec B_*)^2 \Omega \vec C- K \hat z\vec Q_*  \cdot \vec C- i (\vec B_*)^2  K \hat z \times \vec C~~~,
\end{equation}
with $\hat z$ a unit vector along the $z$ axis, and the constraint Eq. \eqref{eq:newchi} becomes an admissability
condition on $\vec C$,
\begin{equation}\label{eq:newchi1}
0=\vec \sigma \cdot \hat z   \times \vec C~~~.
\end{equation}

Writing $F_m$ as a matrix times $C_n$ (and dropping the subscripts $*$, which are implicit from here on) we have
\begin{align}\label{eq:matrix}
F_m=&N_{mn}C_n~~~,\cr
N_{mn}=&(\vec B)^2 \Omega \delta_{mn}- K \delta_{m3} Q_n  - i (\vec B)^2  K \epsilon_{m3n}~~~.\cr
\end{align}
The equation for the characteristics is now
\begin{equation}\label{eq:char}
{\rm det} (N)=0~~~,
\end{equation}
since this is the condition for Eq. \eqref{eq:ceq} to have a solution with nonzero $\vec C$.  However, since evaluation of
the determinant shows that it factorizes into blocks that determine $C_{1,2}$ and a  block that determines $C_3$, a simpler
way to proceed is to work directly from the equations $F_m=0$, which decouple in a corresponding way. Calculating from
Eq. \eqref{eq:ceq},  we find
\begin{align}\label{eq:fcomp}
0=&F_1^{\uparrow,\downarrow}=(\vec B)^2 \big(\Omega C_1^{\uparrow,\,\downarrow}+iK C_2^{\uparrow,\,\downarrow}\big)~~~,\cr
0=&F_2^{\uparrow,\,\downarrow}=(\vec B)^2 \big(\Omega C_2^{\uparrow,\,\downarrow}-iK C_1^{\uparrow,\,\downarrow}\big)~~~,\cr
0=&F_3^{\uparrow,\,\downarrow}=(\vec B)^2 \Omega C_3^{\uparrow,\,\downarrow}- K (\vec Q \cdot \vec C)^{\uparrow,\,\downarrow}~~~,\cr
\end{align}
where $\uparrow,\, \downarrow$ indicate the up and down spinor components, labeled in Eq. \eqref{eq:Psidef} by $\alpha=1,\,2$. Similarly,
the constraint Eq. \eqref{eq:newchi1} becomes $0= -\sigma_1 C_2 +\sigma_2 C_1$, that is
\begin{align}\label{eq:newchi2}
C_2^{\uparrow}=&iC_1^{\uparrow}~~~,\cr
C_2^{\downarrow}=&-i C_1^{\downarrow}~~~,\cr
\end{align}
with no corresponding condition on $C_3^{\uparrow,\,\downarrow}$.
The first two lines of Eq. \eqref{eq:fcomp} together with Eq. \eqref{eq:newchi2} have the solution
\begin{align}\label{eq:c12soln}
C_1^{\uparrow}=&C~,~~C_2^{\uparrow}=iC~,~~\Omega=K~~~,\cr
C_1^{\downarrow}=&C~,~~C_2^{\downarrow}=-iC~,~~\Omega=-K~~~,\cr
\end{align}
with C arbitrary, corresponding to waves with velocity of magnitude $|\Omega/K|=1$.  Thus the modes with $C_{1,2}\neq 0$ are
exactly luminal. Because general background gauge fields are a non-isotropic medium, these modes have nonzero longitudinal
components given by solving the third line of Eq. \eqref{eq:fcomp},
\begin{equation}\label{eq:solveforc3}
C_3=K \big((\vec B)^2 \Omega-K Q_3\big)^{-1} (Q_1 C_1+ Q_2 C_2)~~~.
\end{equation}

The effect on the characteristics of a gauge change $\vec \Psi \to \vec \Psi + \vec D \epsilon$, $\epsilon= E \exp(i \Omega t + i K z)f(t,z)$, where $f$ has a unit slope
discontinuity along the $z$ axis at $x_*$, is to shift $C_3^{\uparrow,\downarrow} \to C_3^{\uparrow,\downarrow} + E^{\uparrow,\downarrow} $,
and thus $ C_3^{\uparrow,\downarrow}$ are gauge degrees of freedom.
In Appendix B, we continue this discussion and show that the longitudinal gauge mode with $C_1=C_2=0, C_3 \neq 0$ also does not propagate superluminally, although in general it is subluminal.

\section{Failure of adiabatic decoupling and inapplicability of the $S$-matrix ``no-go'' theorems}

We show in this section that various ``no-go'' theorems that claim to rule out gauging of higher spin theories
do not apply to the gauged Rarita-Schwinger field.  The reason is that there is a failure of adiabatic decoupling, arising
from the fact that the $\omega$ secondary constraint is homogeneous in the gauge fields.  For a recent paper on
``no-go'' theorems see \cite{mcgady}, which has extensive references to the earlier literature.  In our analysis here
we shall refer specifically to the  paper of Porrati \cite{porrati}, which uses so called ``on-shell'' methods to give limits on
massless high-spin particles.

The analysis of Porrati assumes that ``the general helicity-conserving matrix element of a $U(1)$ current
between on-shell spin $s$ states is $\langle v, p+q|J_{\mu}|u,p\rangle$...'', where $u$ and $v$ are free-space spinors that obey the massless
Dirac equation.  Porrati assumes that the matrix element is bilinear in $u$ and $v$, and ``otherwise depends only on
the momenta''.  We shall see in the following subsections that this assumed form is not realized in the gauged Rarita-Schwinger
theory, where because of the failure of adiabatic decoupling the matrix element in question also depends on
the $U(1)$ gauge field polarization through the dual field-strength $\hat F_{\mu \nu}=\frac{1}{2}\epsilon_{\mu\nu\lambda\sigma} F^{\lambda\sigma}$.  In fact, the initial and final Rarita-Schwinger
spinors both must have a $\hat F_{\mu \nu}$ dependence in order to obey the secondary constraint of Eq. \eqref{eq:constraint2}, and
so the matrix element has the more complicated form $\langle v, p+q, \hat F_{\mu \nu} |J_{\mu}|u,p, \hat F_{\mu \nu}\rangle$.

We show in Sec. 7A that the initial and final Rarita-Schwinger spinors  in the limit of zero gauge field amplitude are equal to free-space spinors $u,v$ of the form assumed by
Porrati,  plus a fermionic gauge transformation that depends explicitly on the photon field strength $\hat F_{\mu \nu}$. This structure arises from the homogeneous form of the secondary constraint, and  corresponds to
an intrinsically non-perturbative aspect of the gauged Rarita-Schwinger
equation.  As another reflection of this, we show in Sec. 7B that one cannot set up a covariant Lippmann-Schwinger equation \cite{lippmann} for the
Rarita-Scwhinger wave function, and so the matrix element that enters into the ``no-go'' theorems does not admit a Born approximation.
In Sec. 7C, we show that a matrix element that has all the required invariances can be formulated using an analog of the
distorted wave Born approximation, in which the initial and final Rarita-Schwinger states have an explicit dependence on the photon
polarizations.

\subsection{The zero amplitude limit of the $\vec\Psi$ equation: retained memory of the gauge field}

As in Sec. 6, let us consider a Rarita-Schwinger field propagating in an external Abelian gauge field.  For convenience, we
assume that the ratio $|\vec E(\vec x)|/|\vec B(\vec x)|\equiv r(\vec x)$ is bounded from above.  In the limit as the
vector potential amplitude $\vec A$ is scaled to zero, Eqs. \eqref{eq:solve1} and \eqref{eq:qdef} become
\begin{align}\label{eq:psi0lim}
\Psi_0(\vec x)=&\vec R(\vec x) \cdot \vec \Psi(\vec x)~~~,\cr
\vec R(\vec x)=&\vec \sigma \cdot \hat B(\vec x) \big(\hat B(\vec x) + r(\vec x)\vec \sigma \times \hat E(\vec x)\big)~~~,\cr
\end{align}
with $\hat B = \vec B /|\vec B|$ and $\hat E=\vec E/|\vec E|$ unit vectors along the $\vec E$ and $\vec B$ fields.  When the
external field is a propagating plane wave with wave vector direction $\hat q$, the unit vectors $\hat q$, $\hat B$ and $\hat E$
form an orthonormal set of constant unit vectors, and $|\vec r(\vec x)|=1$.  We see that because the secondary constraint of Eq. \eqref{eq:constraint2} is homogeneous in the field strengths, the relation between $\Psi_0$ and $\vec \Psi$ retains a memory of the gauge field orientations, and thus of the photon polarization, even in the limit as the field amplitude approaches zero.

In the zero amplitude limit, $D_0=\partial_0$ and $\vec D =\vec \nabla$, so substituting Eq. \eqref{eq:psi0lim} into Eq. \eqref{eq:d0psi3}, the zero amplitude limit for the equation of motion for $\vec \Psi$ becomes
\begin{equation}\label{eq:limd0psi}
\partial_0 \vec \Psi=\vec \nabla  \vec R \cdot\vec \Psi+ i \vec \nabla \times \vec \Psi~~~.
\end{equation}
with the primary constraint now $\vec \sigma \cdot \vec \nabla \times \vec \Psi=0$.
Hence through $\vec R$ the $\vec \Psi$ equation of motion retains a memory of the external fields in the limit of zero amplitude, that is, adiabatic
decoupling has failed.  Let us now consider the situation in which the Rarita-Schwinger field and the external gauge fields are plane waves,  so that   $\vec R$ is a constant and $\vec \Psi$ has the form
\begin{equation}\label{eq:psiwave}
\vec \Psi=\vec C e^{i(\Omega t+\vec k \cdot \vec x)} ~~~.
\end{equation}
Making the fermionic gauge transformation
\begin{align}\label{eq:gaugechange}
\vec \Psi &\to \vec \Psi^{\prime}=\vec \Psi + \vec \nabla \epsilon~~~,\cr
\epsilon =& E e^{i(\Omega t + \vec k \cdot \vec x)}~~~,\cr
\end{align}
$\vec \Psi^{\prime}$ still obeys the zero amplitude primary constraint since $\vec \sigma \cdot \vec \nabla \times \vec \nabla \epsilon=0$.
Then the gauge choice
\begin{equation}\label{eq:echoice}
E=i\frac{\vec R \cdot \vec C}{\vec R \cdot \vec k}
\end{equation}
reduces Eq. \eqref{eq:limd0psi} to the free-space form
\begin{equation}\label{eq:limd0psi1}
\partial_0 \vec \Psi^{\prime}= i \vec \nabla \times \vec \Psi^{\prime}~~~.
\end{equation}
Thus a Rarita-Schwinger plane wave in a zero amplitude gauge field plane wave background is equal to a free-space solution plus
a gauge term that has a memory of the photon polarizations.

\subsection{Breakdown of the Lippmann-Schwinger equation: no Born approximation to scattering}

Let us now examine what happens if one tries to set up  a  covariant Lippmann-Schwinger equation, so as to generate
a Born perturbation series for the Rarita-Schwinger wave function in an external gauge field.  Let us start from
the Rarita-Schwinger equation in the form (see Eq. \eqref{eq:a6})
\begin{equation}\label{eq:gammars}
\gamma^{\eta\nu\rho}D_{\nu}\psi_{\rho}=0~~~.
\end{equation}
Splitting $D_{\nu}$ into $\partial_{\nu}$ and $gA_{\nu}$,
this equation takes the form
\begin{equation}\label{eq:gammars1}
\gamma^{\eta\nu\rho}\partial_{\nu}\psi_{\rho}=-\gamma^{\eta\nu\rho}gA_{\nu}\psi_{\rho}~~~.
\end{equation}
Let us now try to solve this equation as a perturbation series around a free-space
solution by writing
\begin{equation}\label{eq:lippmann1}
\psi_{\rho}(x)=\psi_{\rho}^{\rm free}(x)+ \int d^4y S_{\rho \alpha}(x-y) \gamma^{\alpha\beta\kappa}g A_{\beta}(y) \psi_{\kappa}(y)~~~,
\end{equation}
where $\psi_{\rho}^{\rm free}$ obeys the free-space Rarita-Schwinger equation
\begin{equation}\label{eq:freespace1}
\gamma^{\eta\nu\rho}\partial_{\nu} \psi_{\rho}^{\rm free}=0.
\end{equation}
 If the free-space Green's Rarita-Schwinger Green's function $S_{\rho \alpha}(x-y)$ obeyed
\begin{equation}\label{eq:freegreen1}
\gamma^{\eta\nu\rho}\partial_{x \nu} S_{\rho \alpha}(x-y)=-\delta^{\eta}_{\alpha}\delta^4(x-y)~~~,
\end{equation}
then Eq. \eqref{eq:lippmann1} would reproduce Eq. \eqref{eq:gammars1}.  But in fact the free-space
Green's function cannot obey Eq. \eqref{eq:freegreen1}, because $\partial_{x\eta}\gamma^{\eta\nu\rho}\partial_{x\nu} S_{\rho \alpha}(x-y)=0$;
instead it obeys \cite{freed}
\begin{equation}\label{eq:freegreen2}
\gamma^{\eta\nu\rho}\partial_{x \nu} S_{\rho \alpha}(x-y)=-\delta^{\eta}_{\alpha}\delta^4(x-y)+\partial_{y\alpha}\Omega^{\eta}(x-y)~~~,
\end{equation}
with $\Omega$ necessarily nonvanishing.
Integrating $\partial_{y\alpha}$ by parts onto the factor $\gamma^{\alpha\beta\kappa}g A_{\beta}(y) \psi_{\kappa}(y)$, one
gets
\begin{equation}\label{eq:intbyparts}
\gamma^{\alpha\beta\kappa}gF_{\alpha \beta}(y) \psi_{\kappa}(y)
+\gamma^{\alpha\beta\kappa}g A_{\beta}(y)\partial_{y\alpha} \psi_{\kappa}(y)~~~.
\end{equation}
The first term of this expression vanishes by virtue of the secondary constraint, but the second term is
non-vanishing because the Rarita-Schwinger equation for the exact wave function $\psi_{\kappa}(y)$ is
\begin{equation}\label{eq:intby parts1}
\gamma^{\alpha\beta\kappa}D_{y\alpha} \psi_{\kappa}(y)=0~~~,
\end{equation}
that is, it requires the full covariant derivative $D_{y\alpha}$ in place of its free-space restriction $\partial_{y\alpha}$.
The conclusion from this analysis is that one cannot set up a covariant Lippmann-Schwinger equation for the gauged Rarita-Schwinger
wave function, and thus one cannot develop this wave function into a Born approximation series expansion in powers of the
coupling $g$ to  the external gauge field.

\subsection{Lorentz covariance and mode counting in on-shell Rarita-Schwinger field-photon scattering: a distorted
wave Born approximation analog}
We address finally the question  \cite{witten} of whether one can write down an amplitude for leading order on-shell scattering of Rarita-Schwinger fields from an external electromagnetic field, which  has the requisite relativistic covariance while preserving the correct counting of massless spin $\frac{3}{2}$ propagation modes.  Looking ahead to quantization, an operator effective action for this scattering process can be inferred from the interaction term in Eq. \eqref{eq:action},
\begin{align}\label{eq:effaction}
S_{\rm eff}(\psi_{\mu},A_{\nu}) = &\int d^4x {\cal L}_{\rm eff}(\psi_{\mu},A_{\nu})~~~,\cr
{\cal L}_{\rm eff}(\psi_{\mu},A_{\nu})=& \frac{1}{2}g\,\overline{\psi}_{\mu} (x)
i\epsilon^{\mu\eta\nu\rho}\gamma_5\gamma_{\eta}A_{\nu}(x) \psi_{\rho}(x)~~~,\cr
\end{align}
where we have suppressed spinor indices as in the text from Eq. \eqref{eq:eqmo} onwards.
For Abelian external fields $A_{\nu}$, the covariant derivatives in the
equations of motion and constraints are given by
\begin{equation}\label{eq:covdeviv}
D_{\nu}=\partial_{\nu}+gA_{\nu}~~,~~\overleftarrow{D}_{\nu}=\overleftarrow{\partial}_{\nu}-gA_{\nu}~~~.
\end{equation}
At the outset we shall assume that $A_{\nu}(x)$ is of short range, and vanishes for $|\vec x|>R$ for
some radius $R$.
This effective action, the equations of motion of Eqs. \eqref{eq:eqmo} and \eqref{eq:eqmo1a}, and the primary and secondary
constraints following from them, given in Eqs. \eqref{eq:constraint1} and \eqref{eq:constraint2}, are all relativistically covariant, and so
provide a starting point for calculating a covariant scattering amplitude.
Taking the matrix element of Eq. \eqref{eq:effaction} between an incoming Rarita-Schwinger state  of four-momentum $p$, and an outgoing Rarita-Schwinger
state of four momentum $p'$, we get the corresponding scattering amplitude
\begin{equation}\label{eq:amplitude}
{\cal A}_S = \frac{1}{2}ig\,\int d^4x
 \overline{\psi}_{\mu }(p^{\prime},x)
\epsilon^{\mu\eta\nu\rho}\gamma_5\gamma_{\eta}A_{\nu}(x) \psi_{\rho}(p,x)~~~,
\end{equation}
where $\psi_{\rho}$ and $\overline{\psi}_{\mu}$ are now wave functions, rather than operators, that
obey the Rarita-Schwinger equations of motion in the presence of the external field $A_{\nu}$.

We now introduce source currents for the gauge potential $A_{\nu}$ and the Rarita-Schwinger
wave functions $\psi_{\rho}$ and $\overline{\psi}_{\mu}$, and study their conservation properties.  The source current
to which the gauge potential $A_{\nu}$ couples is  defined by writing the scattering amplitude as
\begin{align}\label{eq:source1}
{\cal A}_S=&\frac{1}{2}ig\,\int d^4x A_{\nu}(x) J^{\nu}(x)~~~,\cr
J^{\nu}(x)=& \overline{\psi}_{\mu }(p^{\prime},x)
\epsilon^{\mu\eta\nu\rho}\gamma_5\gamma_{\eta} \psi_{\rho}(p,x)~~~.\cr
\end{align}
The  source current for the Rarita-Schwinger field $\overline{\psi}_{\mu }(p^{\prime},x)$ is
defined by writing the scattering amplitude as
\begin{align}\label{eq:source2}
{\cal A}_S=&\frac{1}{2}ig\,\int d^4x\overline{\psi}_{\mu }(p^{\prime},x){\cal J}^{\mu}(p,x)~~~,\cr
{\cal J}^{\mu}(p,x)=&\epsilon^{\mu\eta\nu\rho}\gamma_5\gamma_{\eta}A_{\nu}(x) \psi_{\rho}(p,x)~~~.\cr
\end{align}
Finally, the source current  for the Rarita-Schwinger field $ \psi_{\rho}(p,x)$ is
defined by writing the scattering amplitude as
\begin{align}\label{eq:source3}
{\cal A}_S=&\frac{1}{2}ig\,\int d^4x \overline{\cal J}^{\rho}(p^{\prime},x) \psi_{\rho}(p,x)~~~,\cr
\overline{\cal J}^{\rho}(p^{\prime},x)=&\overline{\psi}_{\mu }(p^{\prime},x)\epsilon^{\mu\eta\nu\rho}\gamma_5\gamma_{\eta}A_{\nu}(x) ~~~.\cr
\end{align}

We now show that the three currents that we have just defined are conserved.  For the source current  $J^{\nu}$  for the gauge potential,
we have
\begin{align}\label{eq:cons1}
 \partial_{\nu}J^{\nu}=&
  \overline{\psi}_{\mu }(p^{\prime},x)\overleftarrow{D}_{\nu}
\epsilon^{\mu\eta\nu\rho}\gamma_5\gamma_{\eta} \psi_{\rho}(p,x)\cr
 +& \overline{\psi}_{\mu }(p^{\prime},x)
\epsilon^{\mu\eta\nu\rho}\gamma_5\gamma_{\eta} D_{\nu}\psi_{\rho}(p,x)\cr
=&0~~~,\cr
\end{align}
where the first and second terms on the right vanish by the Rarita-Schwinger equations
for $ \overline{\psi}_{\mu }(p^{\prime},x)$ and $\psi_{\rho}(p,x)$ respectively.
For the source current ${\cal J}^{\mu}(p,x)$ for the spinor $\overline{\psi}_{\mu }(p^{\prime},x)$ , we
have
\begin{align}\label{eq:cons2}
D_{\mu}{\cal J}^{\mu}(p,x)=&\epsilon^{\mu\eta\nu\rho}\gamma_5\gamma_{\eta}\big(\partial_{\mu}A_{\nu}(x)\big) \psi_{\rho}(p,x)\cr
+&\epsilon^{\mu\eta\nu\rho}\gamma_5\gamma_{\eta}A_{\nu}(x) D_{\mu}\psi_{\rho}(p,x)\cr
=&0~~~.\cr
\end{align}
The second term on the right vanishes by the Rarita-Schwinger equation
for  $\psi_{\rho}(p,x)$, while the first term on the right can be rewritten as
\begin{equation}\label{eq:cons2a}
\frac{1}{2}\epsilon^{\mu\eta\nu\rho}\gamma_5\gamma_{\eta}F_{\mu\nu}(x)\psi_{\rho}(p,x)
\end{equation}
and vanishes by the secondary constraint of Eq. \eqref{eq:constraint2}.
Finally, for the source current $\overline{\cal J}^{\rho}(p^{\prime},x)$ for the spinor $ \psi_{\rho}(p,x)$, we have
\begin{align}\label{eq:cons3}
\overline{\cal J}^{\rho}(p^{\prime},x)\overleftarrow{D}_{\rho}=&\overline{\psi}_{\mu }(p^{\prime},x)\epsilon^{\mu\eta\nu\rho}\gamma_5\gamma_{\eta}\big(\partial_{\rho}A_{\nu}(x)\big)\cr
+&\overline{\psi}_{\mu }(p^{\prime},x)\overleftarrow{D}_{\rho}\epsilon^{\mu\eta\nu\rho}\gamma_5\gamma_{\eta}A_{\nu}(x)\cr
=&0~~~.\cr
\end{align}
Again, the second term on the right vanishes by the Rarita-Schwinger equation, while the first term on the right
vanishes by the secondary constraint of Eq. \eqref{eq:constraint2}.

Consider now the following three gauge transformations,
\begin{align}\label{eq:gauge1}
A_{\nu}(x) \to   & A_{\nu}(x) + \partial_{\nu}\Lambda~~~,\cr
\psi_{\rho}(p,x) &\to \psi_{\rho}(p,x)+ D_{\rho} \alpha~~~,\cr
\overline{\psi}_{\mu }(p^{\prime},x) \to & \overline{\psi}_{\mu }(p^{\prime},x) + \overline{\beta}\overleftarrow{D}_{\mu}~~~,\cr
\end{align}
with $\alpha$ and $\beta$ independent spinorial gauge parameters. From Eqs. \eqref{eq:source1}-\eqref{eq:source3},
together with Eqs. \eqref{eq:cons1}-\eqref{eq:cons3},
we find that these transformations each leave the amplitude ${\cal A}_S$ invariant,
\begin{align}
\delta_{\Lambda} {\cal A}_S= & \frac{1}{2}ig\,\int d^4x \big(\partial_{\nu}\Lambda\big) J^{\nu}(x)
=-\frac{1}{2}ig\,\int d^4x \Lambda  \partial_{\nu} J^{\nu}(x)=0~~~,\cr
\delta_{\alpha} {\cal A_S}=&\frac{1}{2}ig\,\int d^4x \overline{\cal J}^{\rho}(p^{\prime},x) D_{\rho} \alpha
=-\frac{1}{2}ig\,\int d^4x \overline{\cal J}^{\rho}(p^{\prime},x)\overleftarrow{D}_{\rho} \alpha=0~~~,\cr
\delta_{\beta} {\cal A}_S=& \frac{1}{2}ig\,\int d^4x    \overline{\beta}\overleftarrow{D}_{\mu}{\cal J}^{\mu}(p,x)
= -\frac{1}{2}ig\,\int d^4x    \overline{\beta}D_{\mu}{\cal J}^{\mu}(p,x)=0~~~.\cr
\end{align}
This, together with the primary and secondary constraints, implies the correct mode-counting for the Rarita-Schwinger
wave functions, since the gauge degrees of freedom do not change the amplitude and so are redundant.

 We next must
specify more precisely the structure of the spinor wave functions entering the formula for ${\cal A}_S$.  Since the
gauge field $A_{\nu}$ is assumed to vanish in the external region $|\vec x|>R$, the Rarita-Schwinger wave functions obey
free field equations in this region.  So for $|\vec x|>>R$ they can be taken asymptotically as plane waves at $t \to \pm \infty$,
\begin{align}\label{eq:extsoln}
\psi_{\mu}(p^{\prime},x)\sim &u_{\mu}(p^{\prime})e^{i p^{\prime} \cdot x}~~,~~t \to +\infty~~~,\cr
\psi_{\rho}(p,x)\sim & u_{\rho}(p) e^{i p \cdot x}~~,~~t \to -\infty          ~~~.\cr
\end{align}
With these boundary conditions, the formula for the amplitude takes the final form
\begin{equation}\label{eq:amplitudefinal}
{\cal A}_S = \frac{1}{2}ig\,\int d^4x
 \overline{\psi}^{(-)}_{\mu }(p^{\prime},x)
\epsilon^{\mu\eta\nu\rho}\gamma_5\gamma_{\eta}A_{\nu}(x) \psi^{(+)}_{\rho}(p,x)~~~.
\end{equation}
The out state (-) and in state (+) boundary conditions used here are analogs of the
boundary conditions used in
the distorted wave Born approximation \cite{dwba}, which the construction of Eq. \eqref{eq:amplitudefinal} resembles.
Equation  \eqref{eq:amplitudefinal} then gives an approximation to the matrix element for
Rarita-Schwinger scattering by the gauge potential.

Rather than invoking the presence of redundant degrees of freedom to count physical Rarita-Schwinger states, we can
follow the usual procedure of imposing a gauge-fixing constraint.  To preserve relativistic and gauge covariance, this
can be taken as the gauge covariant Lorentz gauge condition
\begin{equation}\label{eq:gaugecovlor1}
 \overline{\psi}_{\mu}(p^{\prime},x)\overleftarrow{D}^{\mu}= D^{\rho}\psi_{\rho}(p,x)=0~~~.
 \end{equation}
which is attainable from a generic gauge by the gauge transformation of Eq. \eqref{eq:gaugetrans}, provided that $D^{\mu}D_{\mu}$ is invertible.
In the external region where the gauge field vanishes, one can instead use the condition $\gamma^{\rho}\psi_{\rho}=0$ in place
of the secondary constraint together with the gauge condition $\partial^{\rho}\psi_{\rho}=0$, giving the usual covariant
degree of freedom counting for the incoming and outgoing Rarita-Schwinger wave functions \cite{alvarez}.  Alternatively,
if we are not concerned to maintain manifest Lorentz covariance, we can make a gauge transformation in the external region to
the gauge $\psi_0= \vec \nabla \cdot \vec \psi=0$ used in \cite{dasb}, \cite{freed} to enumerate Rarita-Schwinger degrees of freedom.  When
a non-Lorentz covariant radiation gauge condition is used, scattering matrix elements depend on a unit timelike vector in addition to the
particle momenta, and so the conditions assumed in \cite{porrati} are not obeyed.

Note that if one were to attempt to construct a Born approximation amplitude, in which the Rarita-Schwinger wave functions
in the presence of the gauge field are replaced by plane waves in the interior region where the potential is nonzero, the
arguments given above for compatibility of Lorentz covariance with degree of freedom counting would fail.
The reason for this is that the  spinor source
currents would then no longer be conserved, even to zeroth order in the gauge coupling $g$, because the free particle plane wave solutions
do not obey the secondary constraint of Eq. \eqref{eq:constraint2}.  The non-existence of a satisfactory Born approximation for
Rarita-Schwinger photon scattering agrees with the result obtained in Sec. 7B, that one cannot construct a Lippmann-Schwinger
equation for this process.
To establish compatibility, we have had to use an analog
of the distorted wave Born approximation \cite{dwba}, in which the leading approximation to the amplitude is constructed using interacting rather than free fermion wave functions and does not have a perturbation expansion for  small coupling, $g$ .

When the external Abelian potential is a plane wave field which extends to infinity, there is no large $|\vec x|$ region where the
Rarita-Schwinger solutions reduce to free-space ones. Rather, as shown in Sec. 7A, in the adiabatic decoupling limit of a  zero
amplitude gauge field, the Rarita-Schwinger solutions become free-space solutions plus gauge terms that remember the photon polarization,
and which are necessary to enforce the secondary constraint.  Thus  one cannot attain the kinematic form assumed in the on-shell
``no-go'' theorems.  But as shown here, using distorted Born approximation waves one can write down a consistent
covariant scattering amplitude.

\section{Summary and Remarks}
To conclude, we see that unlike the massive case, the massless gauged Rarita-Schwinger equation leads to a consistent classical theory.  The theory has  the correct counting of propagating non-gauge degrees of freedom with no superluminal wave propagation.  The theory admits a generalized fermionic gauge transformation, and infinitesimal gauge transformations are an invariance of the constrained
flat and curved spacetime actions and of the fermion number. The
gauged Rarita-Schwinger equation has a non-perturbative aspect when the secondary constraint $\omega$ is eliminated, resulting in a breakdown of
adiabatic decoupling, leading to the inapplicability of various $S$-matrix ``no-go''theorems that claim to forbid gauged massless Rarita-Schwinger
fields.  The extension of these results to the quantized
Rarita-Schwinger theory is given in the following paper, where we show that a consistent quantization by the Dirac bracket and path integral
methods is possible, with a manifestly positive semi-definite canonical anticommutator in covariant radiation gauge.
Thus, in the massless case our analysis eliminates the various objections that have been raised to gauging Rarita-Schwinger fields, showing that
non-Abelian gauging of Rarita-Schwinger fields can be contemplated as part of the anomaly cancelation mechanism in constructing grand unified
models.

We conclude with several remarks:

\begin{enumerate}

\item  We have introduced gauge fixing to make  time evolution of the Rarita-Schwinger fields unique, but the analysis of
this paper does not {\it require} gauge fixing.  Specifically, if gauge fixing is not imposed, the correct helicity counting
is still obtained because fermionic gauge degrees of freedom are redundant degrees of freedom, and are not physical.  Gauge
fixing makes this redundancy manifest by providing a condition that excludes the gauge degrees of freedom, but in analogy to
the case of Maxwell electrodynamics, gauge fixing is not needed to get the correct physical state counting.  On the other hand,
in the following paper, where we turn to quantization, gauge fixing is needed.  This can already be anticipated from the form
of the constraint matrix $N$ of Eq. \eqref{eq:detp}, which when gauge fixing is omitted reduces to the single element
${\cal A}= -2ig \vec \sigma \cdot \vec B(\vec x) \delta^3(\vec x-\vec y)$ which is not invertible in the small $\vec B$ limit.
Inversion of the constraint matrix does not enter into the calculations of this paper, but is needed in the following paper
both for Dirac bracket and path integral quantization.

\item A possible exception to the non-perturbative behavior detailed in Sec. 7 is when the $\vec E$ and $\vec B$
gauge fields are random, since if Eq. \eqref{eq:solve1} is replaced by an average, denoted by AV,
\begin{equation}\label{eq:average1}
\langle \Psi_0\rangle_{\rm AV} \simeq \Big\langle  \frac{\vec Q}{(\vec B)^2} \Big\rangle_{\rm AV} \cdot \langle \vec \Psi \rangle_{\rm AV}~~~,
\end{equation}
it becomes
\begin{equation}\label{eq:average2}
\langle \Psi_0\rangle_{\rm AV} \simeq \frac{1}{3}\vec \sigma \cdot \langle \vec \Psi \rangle_{\rm AV}~~~,
\end{equation}
which is compatible with $\langle \Psi_0\rangle_{\rm AV} =\vec \sigma \cdot \langle \vec \Psi \rangle_{\rm AV}=0$, the customary
free Rarita-Schwinger constraints employed in \cite{dasb}, \cite{freed}.  This heuristic observation suggests that Rarita-Schwinger fields coupled to
quantized gauge fields with zero background gauge field  may have a perturbative $g \to 0$ limit.

\item In showing in the Abelian case that there is no superluminal propagation, the inversion of $\vec \sigma \cdot \vec B$ to get
$\Psi_0$ only required $(\vec B)^2 \neq 0$.  In the non-Abelian case, where $\vec B$ is itself a matrix, the conditions for
invertibility are nontrivial and have yet to be analyzed.  We will see in the following paper that this issue is side-stepped when
the constraints are dealt with by the Dirac bracket or path integral procedures, since these do not require inversion of $\vec \sigma \cdot \vec B$ when a  gauge constraint is included.

\end{enumerate}

\section{Acknowledgements}

I wish to thank Edward Witten for  conversations about gauging Rarita-Schwinger fields and Rarita-Schwinger scattering from photons,
among other topics.   I also wish to acknowledge the  various people who
asked about the status of gauged Rarita-Schwinger fields  when I gave seminars on \cite{adler}.  Following on the initial draft
of this paper, I had a fruitful correspondence with Stanley Deser and Andrew Waldron about gauge invariance and counting degrees
of freedom when invariance of the action is conditional on a constraint.   I  wish to thank Thomas Spencer for a very helpful conversation which emphasized the significance of the gauge invariants, and Laurentiu Rodina for an explication of the paper \cite{mcgady} that uses
  ``on-shell'' methods.  This work was supported in part by the National Science Foundation under Grant
No. PHYS-1066293 through the hospitality of the Aspen Center for Physics.

\appendix

\section{Notational conventions and useful identities}

We follow in general the notational conventions of the book {\it Supergravity} by Freedman and Van Proeyen \cite{freed}.
The metric $\eta_{\mu\nu}$ is $(-,+,+,+)$ and the Dirac gamma matrices $\gamma_{\mu}\,,\gamma^{\mu}$ obey the Clifford algebra
\begin{equation}\label{a1}
\gamma_{\mu}\gamma_{\nu}+\gamma_{\nu}\gamma_{\mu}=2\eta_{\mu\nu}~~~.
\end{equation}
They are given in terms of Pauli matrices
$\sigma_j$ by
\begin{align}\label{a2}
\gamma_0=-\gamma^0=&\left( \begin{array} {cc}
 0&-1  \\
 1&0 \\ \end{array} \right)~~~,\cr
 \gamma_j=\gamma^j=&\left( \begin{array} {cc}
 0&\sigma_j  \\
 \sigma_j&0 \\
\end{array}\right) ~~~,\cr
\gamma_5=i\gamma_0\gamma_1\gamma_2\gamma_3=&\left( \begin{array} {cc}
 1&0  \\
 0&-1 \\
\end{array}\right) ~~~.\cr
\end{align}
We also note that
\begin{equation}\label{a3}
\epsilon_{0123}=-\epsilon^{0123}=1~~~,
\end{equation}
the left chiral projector $P_L$ is given by
\begin{equation}\label{a4}
P_L=\frac{1}{2}(1+\gamma_5)~~~,
\end{equation}
and the  spinor $\overline{\psi}$ is defined in terms of the adjoint spinor $\psi^{\dagger}$ by
\begin{equation}\label{a5}
\overline{\psi}=\psi^{\dagger}i\gamma^0~~~.
\end{equation}

As noted in \cite{freed}, the Rarita-Schwinger equation of motion can be written in a number of equivalent forms.
When ordinary derivatives are replaced by gauge covariant derivatives, these are the vector-spinor equations
\begin{align}\label{eq:a6}
\epsilon^{\mu\eta\nu\rho}\gamma_{\eta}D_{\nu}\psi_{\rho}=&0~~~,\cr
\gamma^{\eta \nu \rho} D_{\nu}\psi_{\rho}=&0~~~,\cr
\gamma^{\rho}(D_{\nu}\psi_{\rho}-D_{\rho}\psi_{\nu})=&0~~~,\cr
\gamma^{\alpha}D_{\alpha} (D_{\sigma} \psi_{\nu}-D_{\nu}\psi_{\sigma})=&\gamma^{\rho}\Big([D_{\rho},D_{\sigma}]\psi_{\nu}
+[D_{\nu},D_{\rho}]\psi_{\sigma}+ [D_{\sigma},D_{\nu}]\psi_{\rho}\Big)~~~,\cr
\end{align}
with only the fourth line, which is quadratic in the covariant derivative,  involving more than just a substitution $\partial_{\nu} \to D_{\nu}$ in the formulas
of \cite{freed}. Using $\gamma_{\eta}\gamma^{\eta \nu \rho}=2 \gamma^{\nu \rho}$, these also imply the spinor equation
$\gamma^{\nu\rho} D_{\nu}\psi_{\rho}=0$.
These formulas play a role in verifying stress-energy tensor conservation, as does the
identity \cite{rosen}
\begin{equation} \label{a7}
0=\epsilon^{\lambda\sigma\mu\nu}(A_{\tau}B_{\lambda}C_{\sigma}D_{\mu}E_{\nu}
+ A_{\nu}B_{\tau}C_{\lambda}D_{\sigma}E_{\mu}+ A_{\mu}B_{\nu}C_{\tau}D_{\lambda}E_{\sigma}
+ A_{\sigma}B_{\mu}C_{\nu}D_{\tau}E_{\lambda}+ A_{\lambda}B_{\sigma}C_{\mu}D_{\nu}E_{\tau})~~~,
\end{equation}
with $A_{\tau},\,B_{\lambda},\,C_{\sigma},\,D_{\mu},\,E_{\nu}$ five arbitrary four vectors. This identity
follows from
\begin{equation}\label{a8}
0=\delta_{\tau}^{\alpha}\epsilon^{\lambda\sigma\mu\nu}+\delta_{\tau}^{\nu}\epsilon^{\alpha\lambda\sigma\mu}
+\delta_
{\tau}^{\mu}\epsilon^{\nu\alpha\lambda\sigma}+\delta_{\tau}^{\sigma}\epsilon^{\mu\nu\alpha\lambda}+\delta_{\tau}^{\lambda}\epsilon^{\sigma\mu\nu\alpha}~~~,
\end{equation}
which is easily verified  by noting that $\lambda,\,\sigma,\,\mu,\,\nu$ must take
distinct values from the set $0,1,2,3$, and that $\tau$ must be equal to one of these values.

The fundamental identity for the Pauli matrices is
\begin{equation}\label{a9}
\sigma_a\sigma_b=\delta_{ab}+i\epsilon_{abc} \sigma_c~~~,
\end{equation}
with $\epsilon_{123}=1$ and with the index $c$ summed.
We repeatedly use the following two identities that can be derived from Eq. \eqref{a9}, for a general three vector $\vec A$ that is proportional to a unit
matrix in the spinor space and so commutes
 with $\vec \sigma$,
\begin{align}\label{a10}
\vec \sigma \times (\vec \sigma \times \vec A)=&-2 \vec A+ i \vec \sigma \times \vec A~~~,\cr
(\vec A \times \vec \sigma) \times \vec \sigma=&-2 \vec A+ i \vec A \times \vec \sigma~~~.\cr
\end{align}
Additional useful identities  are
\begin{align}\label{a11}
\vec \sigma \times \vec \sigma =& 2i\vec \sigma ~~~,\cr
\vec \sigma \,\vec \sigma \cdot \vec A=& \vec A-i\vec \sigma \times \vec A ~~~,\cr
\vec \sigma \cdot \vec A \, \vec \sigma =& \vec A+i\vec \sigma \times \vec A ~~~, \cr
(\vec \sigma \times \vec A) \cdot \vec \sigma=&-2i \vec \sigma \cdot \vec A~~~,\cr
\vec \sigma \cdot (\vec \sigma \times \vec A)=& 2i \vec \sigma \cdot \vec A~~~,\cr
\sigma_a\sigma_b=&2\big(\delta_{ab}-\frac{1}{2}\sigma_b\sigma_a\big)~~~,\cr
\vec B=i\vec A -\vec A \times \vec \sigma  \leftrightarrow & \vec A=\frac{1}{2}(\vec B\times \vec \sigma)~~~.\cr
\end{align}

Gauge field covariant derivatives are
\begin{equation}\label{a12}
D_{\mu}=\partial_{\mu}+gA_{\mu}~~~,
\end{equation}
with the gauge potential $A_{\mu}=A_{\mu}^A t_A $ and the gauge generators $t_A$ anti-self-adjoint,
and with the components $A_{\mu}^A$ self-adjoint.  The non-Abelian generators $t_A$ obey the compact  Lie algebra
\begin{equation}\label{a13}
[t_A,t_B]=f_{ABC}t_C~~~;
\end{equation}
in the Abelian case we replace $t_A$ by $-i$.  In writing field strengths $\vec E$ and $\vec B$ we pull out
an additional factor of $i$ to make them self-adjoint, so that we have the identities
\begin{align}\label{a14}
\vec D \times \vec D =& -i g \vec B~~~,\cr
[\vec D, D_0] =& -i g \vec E~~~.\cr
\end{align}
We will also write a right-acting three-vector covariant derivative as $\overrightarrow  D=\overrightarrow \nabla +g \vec A$, and define a left-acting three-vector covariant derivative as $\overleftarrow D=\overleftarrow \nabla-g\vec A$, so that we
have the integration by parts formulas
\begin{align}\label{a15}
\int d^3x A \overrightarrow  D_{\vec x} B=&-\int d^3x A  \overleftarrow D_{\vec x} B~~~,\cr
\vec{D}_{\vec x} \delta^3(\vec x -\vec y)=&-\delta^3(\vec x -\vec y)\overleftarrow{D}_{\vec y}~~~.\cr
\end{align}
An analogous definition is used for the operators $\vec L$ and $\overleftarrow{L}$ which enter the gauge fixing condition.

At the classical level, variables will be either Grassmann even or odd.  Irrespective of the Grassmann parity of monomials
$A$ and $B$, the adjoint operation is defined by \cite{freed}
\begin{equation}\label{a16}
(AB)^{\dagger}=B^{\dagger}A^{\dagger}~~~.
\end{equation}
For classical brackets, we follow the convention of Henneaux and Teitelboim \cite{teitel},
\begin{equation}\label{a17}
[F,G]_C=\left(\frac{\partial F}{\partial q^i}\frac{\partial G}{\partial p_i}-
\frac{\partial F}{\partial p_i}\frac{\partial G}{\partial q^i}\right)+(-)^{\epsilon_F}
\left(\frac{\partial^L F}{\partial \theta^{\alpha} }\frac{\partial^L G}{\partial \pi_{\alpha}}+
\frac{\partial^L F}{\partial \pi_{\alpha}}\frac{\partial^L G}{\partial \theta^{\alpha}}\right)~~~,
\end{equation}
with $\epsilon_F$ the Grassmann parity of $F$, with $\partial^L$ a Grassmann derivative acting from the left, and with $q^i,\,p_i$ ($\theta^{\alpha},\,\pi_{\alpha}$) canonical coordinates
and momenta of even (odd) Grassmann parity. Using the classical bracket, the Dirac  bracket is constructed  from
the constraints as discussed in Sec. 2 of the following paper.   To make the transition to quantum theory, the quantum commutator (anticommutator) is defined to be $i\hbar$ times the corresponding Dirac bracket (with $\hbar=1$ in our notation).   Classical canonical brackets are always denoted, as above, by
a subscript $C$, with a subscript $D$ used for the corresponding Dirac bracket. We use the standard notations $[A,B]=AB-BA$ for the  commutator and $\{A,B\}=AB+BA$ for the anticommutator.

To calculate the Dirac bracket, we  use block inversion of a matrix.  Let
\begin{align}\label{a18}
M=&\left( \begin{array} {cc} A_1  & A_2 \\A_3 &A_4\\ \end{array} \right)~~~,\cr
M^{-1}=&\left( \begin{array} {cc} B_1  & B_2 \\B_3 &B_4\\ \end{array} \right)~~~,\cr
\end{align}
with $A_1,...,A_4$ themselves matrices.  Then when $A_4$ is non-singular, the blocks $B_1,...,B_4$ of $M^{-1}$  are given by
\begin{align}\label{a19}
\Delta\equiv& A_1-A_2 A_4^{-1} A_3 ~~~,\cr
B_1=& \Delta^{-1}~~~,\cr
B_2=&-\Delta^{-1}A_2 A_4^{-1}~~~,\cr
B_3=&-A_4^{-1}A_3 \Delta^{-1}~~~,\cr
B_4=&A_4^{-1}+A_4^{-1} A_3 \Delta^{-1}A_2 A_4^{-1}~~~.\cr
\end{align}
Even though the blocks are noncommutative, Eqs. \eqref{a18} and \eqref{a19} give an inverse that obeys $M^{-1}M=MM^{-1}=1$.

When the constraints $\phi_a$ and $\chi_a$ are combined into an 8 element set of constraints $\kappa_a=\phi_a,\,\kappa_{a+4}=\chi_a,\,a=1,...,4$
then the bracket matrix $S_{ab}(\vec x, \vec y)\equiv [\kappa_a(\vec x),\kappa_b(\vec y)]_C $ can be expressed in terms of the matrix $M_{ab}(\vec x,\vec y)$ of Eq. \eqref{eq:nonvanishbracks} as
\begin{equation}\label{a20}
S(\vec x,\vec y)=\left( \begin{array} {cc}
 0&M(\vec x,\vec y)  \\
 M^T(\vec y,\vec x)&0 \\ \end{array} \right)
 ~~~~,
\end{equation}
where $M_{ab}^T(\vec x, \vec y)=M_{ba}(\vec x, \vec y)$ is the matrix transpose.  Defining the inverse $M^{-1}(\vec x,\vec y)$ that obeys
$\int d^3z M^{-1}(\vec x,\vec z )M(\vec z, \vec y)=\int d^3z M(\vec x,\vec z) M^{-1}(\vec z,\vec y)= \delta^3(\vec x -\vec y)$, it is easy to verify that
\begin{equation}\label{a21}
S^{-1}(\vec x,\vec y)=\left( \begin{array} {cc}
 0&M^{T\,-1}(\vec y,\vec x)  \\
 M^{-1}(\vec x,\vec y)&0 \\ \end{array} \right)
 ~~~~.
\end{equation}

\section{Analysis of the Rarita-Schwinger field in an external Abelian gauge field:
propagation of the longitudinal gauge mode}

We continue here the analysis begun in Sec. 5  to study  propagation of the longitudinal gauge mode.
 We must now solve for $C_3^{\uparrow,\,\downarrow}$ starting from Eq.
\eqref{eq:fcomp}  with $C_{1,2}=0$,   so the third line
of Eq. \eqref{eq:fcomp} simplifies to
\begin{align}\label{eq:fcomp3}
0=&(\vec B)^2 \Omega C_3^{\uparrow,\,\downarrow}- K ( Q_3   C_3)^{\uparrow,\,\downarrow}~~~,\cr
Q_3=&B_1E_2-B_2E_1 + B_3 \vec \sigma \cdot (\vec B+i\vec E)-i\vec B\cdot \vec E \sigma_3~~~.\cr
\end{align}
Writing this as
\begin{equation}\label{eq:nmatrixdef}
\left( \begin{array} {c} 0 \\0 \\ \end{array} \right)
=\left( \begin{array} {cc} U_{11} & U_{12} \\U_{21} & U_{22} \\ \end{array} \right)
\left( \begin{array} {c} C_3^{\uparrow} \\ C_3^{\downarrow} \\ \end{array} \right)~~~,
\end{equation}
we find for the matrix elements
\begin{align}\label{eq:matrixelts}
U_{11}=&(\vec B)^2 \Omega-K[B_1E_2-B_2E_1-i(B_1E_1+B_2E_2)+B_3^2]~~~,\cr
U_{22}=&(\vec B)^2 \Omega-K[B_1E_2-B_2E_1+i(B_1E_1+B_2E_2)-B_3^2]~~~,\cr
U_{12}=&-KB_3[B_1+iE_1-i(B_2+iE_2)]~~~,\cr
U_{21}=&-KB_3[B_1+iE_1+i(B_2+iE_2)]~~~.\cr
\end{align}
The equation $0=\rm{det}(U)=U_{11}U_{22}-U_{12}U_{21}$ reduces, after dividing by an overall
factor of $(\vec B)^2$, to
\begin{equation}\label{eq:omegakeq}
0=(\vec B)^2 \Omega^2 - 2 \Omega K (B_1 E_2-B_2 E_1) + K^2 (E_1^2+E_2^2-B_3^2)~~~,
\end{equation}
with the solution
\begin{align}\label{eq:omegaksoln}
\frac{\Omega}{K}=&\frac{X \pm Y^{1/2}}{(\vec B)^2}~~~,\cr
X=& B_1E_2-B_2E_1~~~,\cr
Y=& (B_1E_2-B_2E_1)^2-(\vec B)^2(E_1^2+E_2^2-B_3^2) ~~~.\cr
\end{align}

The analysis of the solutions of Eqs. \eqref{eq:omegakeq} and \eqref{eq:omegaksoln} divides into two cases, according to whether
the roots of Eq. \eqref{eq:omegaksoln} are both real, or both complex. The roots are both complex if
\begin{equation}\label{eq:complex}
(B_1E_2-B_2E_1)^2<(\vec B)^2(E_1^2+E_2^2-B_3^2)~~~,
\end{equation}
which can be rearranged algebraically to the form
\begin{equation}\label{eq:complex1}
[(\vec B)^2-(E_1^2+E_2^2)]B_3^2<(B_1^2+B_2^2)(E_1^2+E_2^2)\cos^2\phi~~~,
\end{equation}
where we have written
\begin{align}\label{eq:phidef}
B_1E_2-B_2E_1=&(B_1^2+B_2^2)^{1/2}(E_1^2+E_2^2)^{1/2} \sin\phi~~~,\cr
B_1E_1+B_2E_2=&(B_1^2+B_2^2)^{1/2}(E_1^2+E_2^2)^{1/2} \cos\phi~~~.\cr
\end{align}
Since the right hand side of Eq. \eqref{eq:complex1} is non-negative, when the
left hand side is negative the inequality is satisfied, and both roots are complex.
Hence a necessary (but not sufficient) condition for both roots to be real is
\begin{equation}\label{eq:realcond}
(\vec B)^2-(E_1^2+E_2^2)>0~~~.
\end{equation}

\subsection{The hyperbolic case: both roots real}

When both roots are real, Eq. \eqref{eq:fcomp3} describes the hyperbolic case of propagating waves.  Introducing
the velocity $V=\Omega/K$, Eq. \eqref{eq:omegakeq} can be written as
\begin{equation}\label{eq:omegakeqnew}
0=(\vec B)^2 V^2-2V(B_1E_2-B_2E_1)+E_1^2+E_2^2-B_3^2~~~,
\end{equation}
which can be rearranged algebraically to the form
\begin{equation}\label{eq:omegakeqnew1}
[(B_1^2+B_2^2)^{1/2}-(E_1^2+E_2^2)^{1/2}]^2+(\vec B)^2(V^2-1)=2(B_1^2+B_2^2)^{1/2}(E_1^2+E_2^2)^{1/2}(V \sin\phi -1)~~~.
\end{equation}
Let us now assume that $V^2>1$, and show that this leads to a contradiction.  When $V^2>1$, the left hand side of
Eq. \eqref{eq:omegakeqnew1} is nonnegative, which implies that $V \sin\phi$ on the right must be nonnegative, and
so can be replaced by its absolute value.  Hence the right hand side of Eq. \eqref{eq:omegakeqnew1} obeys the
inequality
\begin{equation}\label{eq:ineq1}
2(B_1^2+B_2^2)^{1/2}(E_1^2+E_2^2)^{1/2}(V \sin\phi -1)=2(B_1^2+B_2^2)^{1/2}(E_1^2+E_2^2)^{1/2}(|V \sin\phi| -1)
\leq 2 (\vec B)^2 (|V| -1)~~~,
\end{equation}
where we have used Eq. \eqref{eq:realcond}.  But the left hand side of  Eq. \eqref{eq:omegakeqnew1} obeys the inequality
\begin{equation}\label{eq:ineq2}
[(B_1^2+B_2^2)^{1/2}-(E_1^2+E_2^2)^{1/2}]^2+(\vec B)^2(V^2-1)\geq (\vec B)^2(|V|+1) (|V|-1)>2(\vec B)^2(|V|-1)~~~,
\end{equation}
which is a contradiction, since a real number cannot be strictly less than itself.  Hence we must have $V^2 \leq 1$, and
there is no superluminal propagation.

\subsection{The elliptic case: both roots complex}

When both roots are complex, Eq. \eqref{eq:fcomp3} describes the elliptic case in which  there
are no propagating waves; when a propagating wave enters an elliptic region from a hyperbolic one it will be damped to zero
amplitude.  However, in the case of weak damping one can still define a wave velocity and ask what its magnitude is.
When both roots are imaginary, Eq. \eqref{eq:omegaksoln} takes the form
\begin{align}\label{eq:omegaksoln1}
\frac{\Omega}{K}=&\frac{X \pm i(-Y)^{1/2}}{(\vec B)^2}~~~,\cr
X=& B_1E_2-B_2E_1~~~,\cr
-Y=& -(B_1E_2-B_2E_1)^2+(\vec B)^2(E_1^2+E_2^2-B_3^2) ~~~.\cr
\end{align}
Regarding $\Omega$ as real  and the wave number $K$ as complex, the effective propagation velocity has the magnitude
\begin{equation}\label{eq:effvel}
|V_{\rm eff}|=\Big|\frac{\Omega}{K_R}\Big|=\frac{X^2-Y}{(\vec B)^2 |X|}=\frac{E_1^2+E_2^2-B_3^2}{|B_1E_2-B_2E_1|}~~~.
\end{equation}
The condition for weak damping is $-Y<<X^2$, which can be rewritten as
\begin{equation}\label{eq:weakdamp}
(\vec B)^2 (E_1^2+E_2^2-B_3^2)<<2 (B_1E_2-B_2E_1)^2~~~,
\end{equation}
and implies
\begin{equation}\label{eq:weakdamp1}
|V_{\rm eff}|<<\frac{2|B_1E_2-B_2E_1|}{(\vec B)^2}\leq \frac{2 |\Vec E|}{|\vec B|}~~~.
\end{equation}
Hence as long as $2|\vec E|$ is not much larger than $|\vec B|$, which is required by the vacuum stability condition $|\vec E| < |\vec B|$, the damped wave propagation velocity is subluminal.


\begin{thebibliography}{99}

\bibitem{marcus}  N. Marcus, {\it Phys. Lett.} B {\bf 157}, 383 (1985).
\bibitem{adler} S. L. Adler, {\it Int. J. Mod. Phys.} A {\bf 29}, 1450130 (2014).
\bibitem{johnson} K. Johnson and E. C. G. Sudarshan, {\it Ann. Phys.} {\bf 13}, 126 (1961).
\bibitem{velo} G. Velo and D. Zwanziger, {\it Phys. Rev.} {\bf 186}, 1337 (1969)
\bibitem{hortacsu} M. Hortacsu, {\it Phys. Rev.} D {\bf 9}, 928 (1974).
\bibitem{deser} S. Deser and A. Waldron, {\it Nucl. Phys.} B {\bf 631}, 369 (2002).
\bibitem{dasa}  D. Z. Freedman and A. Das,  {\it Nucl. Phys.} B {\bf 120}, 221 (1977).
\bibitem{grisaru}  M. T. Grisaru, H. N. Pendleton, and P. van Nieuwenhuizen, {\it Phys. Rev.} D {\bf 15}, 996 (1977);
M. T. Grisaru and H. N. Pendleton, {\it Phys. Lett.} B {\bf 67}, 323 (1977).
\bibitem{porrati}  M. Porrati, Phys. Rev. D 78, 065016 (2008).
\bibitem{mcgady}  D. A. McGady and L. Rodina, {\it Phys. Rev.} D {\bf 90}, 084048 (2014).
\bibitem{dasb} A. Das and D. Z. Freedman, {\it Nucl. Phys.} B {\bf 114}, 271 (1976).
\bibitem{das1}  A. Das, {\it Phys. Rev.} D {\bf 18}, 2065 (1978).
\bibitem{allcock} G. R. Allcock and S. F. Hall, {\it J. Phys. A: Math. Gen.} {\bf 10}, 267 (1977).
\bibitem{dirac} P. A. M. Dirac, {\it Proc. Roy. Soc.} A {\bf 246}, 326 (1958).
\bibitem{schwinger} J. Schwinger, {\it Phys. Rev.} {\bf 82}, 664 (1951), Sec. VI.
\bibitem{courant} R. Courant and D. Hilbert, {\it Methods of Mathematical Physics}, Vol. 2, Wiley-Interscience (1962), Ch. VI.
\bibitem{madore} J. Madore and W. Tait, {\it Commun. math. Phys.} {\bf 30}, 201 (1973); J. Madore, {\it Phys. Lett.} B {\bf 55},
217 (1975).
\bibitem{lippmann}  B. A. Lippmann and J. Schwinger, {\it Phys. Rev. Lett.} {\bf 79}, 469 (1950).
\bibitem{witten} E. Witten, private communication.
\bibitem{dwba} N. Austern, {\it Direct Nuclear Reaction Theories}, Wiley-Interscience, New York (1970), Eq. (4.51) p. 79;
G. R. Satchler, {\it Introduction to Nuclear Reactions}, Oxford University Press, New York (1990), Eq. (4.57) p. 206,
Eq. (4.58) p. 207.
\bibitem{alvarez}  L. Alvarez-Gaum\'e and E. Witten, Nucl. Phys. B 234, 269 (1983), Sec. 7.
\bibitem{freed} D. Z. Freedman and A. Van Proeyen, {\it Supergravity}, Cambridge University Press (2012), flyleaf and
Secs. 2.1--2.3.
\bibitem{rosen} L. Rosenberg, {\it Phys. Rev.} {\bf 129}, 2786 (1963).
\bibitem{teitel} M. Henneaux and C. Teitelboim, {\it Quantization of Gauge Systems}, Princeton University Press (1992),
pp. 146 and 273.

\end{thebibliography}
\end{document}